\documentclass[journal=jctcce,manuscript=article]{achemso}

\usepackage{chemformula} 
\usepackage[T1]{fontenc} 
\usepackage{amsmath}
\usepackage{amssymb}
\usepackage{gensymb}
\usepackage{bm}
\usepackage{graphicx}
\usepackage{float}
\usepackage[colorlinks=true, allcolors=blue]{hyperref}
\usepackage[dvipsnames]{xcolor}
\usepackage{booktabs}
\usepackage[capitalize]{cleveref}
\usepackage{relsize}
\usepackage{xspace}
\usepackage{multirow}
\usepackage{chemformula}
\usepackage{gensymb}
\usepackage[dvipsnames]{xcolor}
\usepackage{marginnote}

\author{Francesco De Roma}
\affiliation{DISAT, Institute of Chemical Engineering, Politecnico di Torino, C.so Duca degli Abruzzi 24, Torino 10129, Italy}
\author{Luca Maffioli}
\affiliation{Department of Mathematics, School of Science, Computing and Engineering Technologies, Swinburne University of Technology, P.O. Box 218, Hawthorn 3122, Victoria, Australia}
\author{Edward R. Smith}
\affiliation{Department of Mechanical and Aerospace Engineering, Brunel University London, Uxbridge UB8 3PH, UK}
\email{edward.smith@brunel.ac.uk}
\author{Antonio Buffo}
\email{antonio.buffo@polito.it}
\affiliation{DISAT, Institute of Chemical Engineering, Politecnico di Torino, C.so Duca degli Abruzzi 24, Torino 10129, Italy}

\title{Study of arbitrarily low shear rate rheology using dissipative particle dynamics}

\abbreviations{IR,NMR,UV}

\begin{document}

\singlespace
\begin{abstract}
    The use of dissipative particle dynamics (DPD) simulation to study the rheology of fluids under shear has always been of great interest to the research community.
    Despite being a powerful tool, a limitation of DPD is the need to use high shear rates to obtain viscosity results with a sufficiently high signal-to-noise ratio (SNR).
    This often leads to simulations with unrealistically large deformations that do not reflect typical stress conditions on the fluid.
    In this work, the transient time correlation function (TTCF) technique is used for a simple Newtonian DPD fluid to achieve high SNR results even at arbitrarily low shear rates.
    The applicability of the TTCF on DPD systems is assessed, and the modifications required by the nature of the DPD force field are discussed.
    The results showed that the standard error (SE) of viscosity values obtained with TTCF is consistently lower than that of the classic averaging procedure across all tested shear rates.
    Moreover, the SE resulted proportional to the shear rate, leading to a constant SNR that does not decrease at lower shear rates.
    Additionally, the effect of trajectory mapping on DPD is studied, and a TTCF approach that does not require mappings is consolidated.
    Remarkably, the absence of mappings has not reduced the precision of the method compared with the more common mapped approach.
\end{abstract}

\section{Introduction}
\label{sec:introduction}

The study of the rheology of fluids has always been a topic of interest both for industrial applications and for theoretical understanding.
Almost all industrial processes involve flowing fluids, many of which have a complex rheology, which depends on multiple factors, such as temperature, shear rate, and composition.
Deeper knowledge of the relationship between these variables and the viscosity of the fluid is highly beneficial for process design and optimization.
At the same time, this knowledge can help the theoretical understanding behind the stress response of the fluids, which is largely dependent on phenomena occurring at smaller scales than the macroscopic one.
The popularity of modeling and atomistic simulations has therefore increased over time, driven by the desire to link the microscopic picture and the macroscale rheological behavior of fluids.

Among the many microscopic methods, dissipative particle dynamics (DPD) has attracted attention due to its coarse-grained description of molecules.
This approach reduces the computational resources needed to simulate a system, with respect to a full atomistic description, by simplifying only partially the chemical specificity of the model.
DPD has proven to be an effective technique for simulating complex fluids such as polymer solutions \cite{droghetti2018dissipative,lauriello2025interpretable}, interfacial systems \cite{ferrari2022molecular,ferrari2023application}, and surfactant solutions in water \cite{anderson2018micelle,panoukidou2019constructing}.
The results achieved for the equilibrium structural properties by previous works \cite{wand2020relationship,delregno2021critical} increased interest in the use of the technique to predict the transport properties of fluids.

Since DPD preserves hydrodynamics, its ability to predict transport properties has been a focus since its initial use in studies \cite{groot1997dissipative}.
Over the years, the transport properties of DPD fluids have been investigated from both theoretical \cite{marsh1997static} and computational points of view.
Recently, the rheology of simple DPD fluids has been studied using equilibrium and non-equilibrium methods \cite{boromand2015viscosity,lauriello2021simulation,lauriello2023development}.

The Green-Kubo \cite{green1954markoff,kubo1957statistical} and Einstein-Helfand \cite{helfand1960transport} relations are used to estimate the zero shear viscosity, and versions specifically modified for DPD have been proposed for the first \cite{ernst2005generalized,jung2016computing} and the second \cite{panoukidou2021comparison} methods.
When studying the rheology of many realistic fluids, it is more interesting to analyze the shear-dependent behavior, which requires a non-equilibrium method.
This is the case of complex fluids, whose rheology has been explored by different works \cite{droghetti2018dissipative,nafarsefiddashti2020highfidelity,hendrikse2023anionic,deroma2024application} with satisfactory results, while still exhibiting some limitations.
Moreover, these works show that the viscosity at low shear rates is highly uncertain \cite{boromand2015viscosity,hendrikse2023anionic,deroma2024application}, requiring the use of high shear rates.
It is also documented that high shear rates can cause unexpected results, such as shear thickening in simple fluids \cite{boromand2015viscosity,droghetti2018dissipative}, or the disruption of microstructures in complex fluids \cite{hendrikse2023anionic}.

Being a coarse-grained model, DPD is based on a set of reduced units of measurement and requires conversion factors to recover values in meaningful physical units.
The derivation of conversion factors is well established for equilibrium simulations \cite{lauriello2021simulation}, and depends on the characteristics of the system. 
However, using the same approach in non-equilibrium simulations could produce non-physical results, \textit{e.g.} unrealistically high shear rates \cite{droghetti2018dissipative}.
Consequently, the application of high shear rates to DPD simulations often requires the development of \textit{ad hoc} conversion factors \cite{deroma2024application}, which are not always related to the characteristics of the system.

As with DPD, molecular dynamics (MD) simulations also require high shear rates to obtain a sufficiently high signal-to-noise ratio (SNR) \cite{todd2017nonequilibrium,ewen2018advances}.
Converted to real units, these shear rates are many orders of magnitude higher than those that can be applied experimentally.
To tackle this issue, the transient-time correlation function (TTCF) formalism \cite{evans1990statistical,evans2016fundamentals} has been applied to MD simulations to improve the signal-to-noise ratio.
TTCF is a non-linear generalization of the Green-Kubo formula \cite{green1954markoff,kubo1957statistical}, which is based on the evaluation of the transient response of a system after the imposition of an external field.
Systems under many types of different external fields have been studied, with the common denominator of using external fields of strength that can be compared with experimental conditions.
For tribological applications, the behavior of monoatomic \cite{delhommelle2005simulation,bernardi2012response,maffioli2022slip} and molecular fluids \cite{bernardi2016local} between sliding solid surfaces has been investigated.
The TTCF method is not limited to simple shear, and the condition of elongational flow \cite{todd1997application} and mixed flows \cite{hartkamp2012transienttime} have been studied.
 Consequently, the method is suitable for the study of the rheology of monoatomic fluids \cite{morriss1987application,evans1988transienttimecorrelation,borzsak2002shear,desgranges2009accurate}, liquid metals \cite{desgranges2008shear,desgranges2008molecular}, as well as molecular fluids \cite{pan2006prediction,mazyar2009transient}.

In this work, the applicability of the TTCF formalism to a DPD system is investigated to show the adaptations required by the peculiarities of the DPD force field.
The objective is to standardize the use of TTCF for studying the shear-dependent rheology of DPD fluids under the of simple shear flow.
The results presented in this work refer to a simple DPD fluid to demonstrate the effectiveness of the method in a controlled Newtonian case.
The main advantage of using TTCF is the possibility to simulate the fluid system at very low shear rates, while retaining a high signal-to-noise ratio.
In this way, it will be possible to link the DPD conversion factors to the characteristics of the system and to obtain results that can be compared with experimental data.
A successful method for simulating low shear rates is expected to yield more insughtful results when applied to complex fluids.
In such systems, a weaker external field is expected to produce a more realistic deformation of the microstructures.

The paper is organized as follows: the systems studied and the methods used are described in \Cref{sec:methods_and_computational_details}, together with the modifications implemented to the TTCF approach and the computational details, while in \Cref{sec:results_and_discussion} the results obtained from the simulations are presented and discussed.
Eventually, the conclusions of this work are illustrated \Cref{sec:conclusions}.

\section{Methods and computational details}
\label{sec:methods_and_computational_details}

\subsection{Studied systems}
\label{subsec:studied_systems}

The present work focuses on the application of TTCF to Dissipative Particle Dynamics (DPD), with a Lennard-Jones (LJ) fluid used as a benchmark.
Since the DPD model requires some modification to the standard TTCF approach (see \Cref{subsubsec:dpd_and_mappings}), the LJ fluid, which has been extensively studied, is used to test the correctness of the implementation and eliminate potential coding errors. 

\subsubsection{Lennard-Jones model}
\label{subsubsec:lj_fluid}

A truncated and shifted Lennard-Jones potential is used to model the fluid, resulting in a Weeks-Chandler-Anderson (WCA) potential \cite{weeks1971role}:
\begin{equation}\label{eq:lj_potential}
    \phi(r_{ij}) = 
    \begin{cases}
        4\varepsilon \left[\left(\frac{\sigma}{r_{ij}}\right)^{12} - \left(\frac{\sigma}{r_{ij}}\right)^{6}\right] + \phi_c, & \text{if } r_{ij} \leq r_c \\
        0, & \text{if } r_{ij} > r_c
    \end{cases}
\end{equation}
where $r_{ij}$ is the distance between particles $i$ and $j$, $\sigma$ is the particle diameter, $\varepsilon$ is the potential well, and $r_c$ is the cutoff radius.
The potential is truncated at $r_c = 2^{1/6}\sigma$, and $\phi_c$ is the constant that shifts the potential to ensure the continuity of the function at the cutoff radius.
To maintain consistency with previous works \cite{borzsak2002shear,maffioli2024ttcf4lammps}, the system is studied at the Lennard-Jones triple point, with a reduced density $\rho^* = \rho\sigma^3 = 0.8442$ and a reduced temperature $T^* = k_B T/\varepsilon = 0.722$, with $k_B$ being the Boltzmann constant.
Proceeding in this way, the results can be compared with those obtained by previous works, and the implementation can be tested against a well-known system.

\subsubsection{DPD simple fluid model}
\label{subsubsec:dpd_simple_fluid_model}

Dissipative Particle Dynamics is a computational method that relies on a coarse-grained description of the molecules, which are grouped in larger particles, called beads.
After its first introduction by \citeauthor{koelman1993dynamic}, it was further developed by \citeauthor{groot1997dissipative}, while \citeauthor{espanol1995statistical} studied its formalization from a statistical mechanics point of view.
The coarse-graining (CG) approach allows mesoscopic systems to be studied, enlarges the available spatial and temporal scales, and requires fewer computational resources than all-atom molecular dynamics (MD).
According to the standard DPD model, two beads $i$ and $j$ interact through three pairwise forces, the conservative force $\boldsymbol{F}^C_{ij}$, the dissipative force $\boldsymbol{F}^D_{ij}$, and the random force $\boldsymbol{F}^R_{ij}$.
The conservative force is a soft repulsive force that allows the beads to overlap and has the following functional form:
\begin{equation}\label{eq:dpd_conservative_force}
    \boldsymbol{F}^C_{ij} = \begin{cases}
        a_{ij} \left(1 - \cfrac{r_{ij}}{r_c}\right) \hat{\boldsymbol{r}}_{ij}, & r_{ij} \leq r_c \\
        0, & r_{ij} > r_c
    \end{cases}
\end{equation}
where $a_{ij}$ is the repulsion parameter, which depends on the type of beads, $r_{ij} = |\boldsymbol{r}_{ij}| = |\boldsymbol{r}_i - \boldsymbol{r}_j|$ is the distance between the beads, $\hat{\boldsymbol{r}}_{ij} = \boldsymbol{r}_{ij}/r_{ij}$ is the unit vector pointing from bead $j$ to bead $i$, and $r_c$ is the cutoff radius.
The dissipative and random forces are instead defined as:
\begin{equation}\label{eq:dpd_dissipative_force}
    \boldsymbol{F}^D_{ij} = -\gamma w^D(r_{ij})(\boldsymbol{r}_{ij}\cdot\boldsymbol{v}_{ij})\hat{\boldsymbol{r}}_{ij}
\end{equation}
\begin{equation}\label{eq:dpd_random_force}
    \boldsymbol{F}^R_{ij} = \sigma w^R(r_{ij})\frac{\xi_{ij}}{\sqrt{\Delta t}}\hat{\boldsymbol{r}}_{ij}
\end{equation}
where $w^D(r_{ij})$ and $w^R(r_{ij})$ are their respective weight functions, $\boldsymbol{v}_{ij} = \boldsymbol{v}_i - \boldsymbol{v}_j$ is the relative velocity between the two beads, and $\xi_{ij}$ is a random number drawn from a Gaussian distribution with zero mean and unit variance.
The dissipative force causes a decrease of the energy in the system, which is restored by the random force.
Consequently, these two forces can act as a thermostat and \citeauthor{espanol1995statistical} described the relative magnitude required for these forces to respect the fluctuation-dissipation theorem.
The dissipative parameter $\gamma$ and the random parameter $\sigma$ must be related to each other using the following equation
\begin{equation}\label{eq:gamma_sigma_relation}
    \sigma^2 = 2\gamma k_BT,
\end{equation}
where $k_B$ is the Boltzmann constant and $T$ is the temperature of the system.
In addition to this, the following relation between the weight functions must be enforced:
\begin{equation}\label{eq:weight_function_realation}
    w^D(r_{ij}) = \left[w^R(r_{ij})\right]^2.
\end{equation}
Under these conditions, the energy in the system is conserved and the correct temperature is ensured, as in an NVT ensemble.
In this work, as is common in the literature, the weight for the random force was chosen equal to the functional form of the conservative force, hence:
\begin{equation}\label{eq:dpd_weights}
    w^D(r_{ij}) = \left[w^R(r_{ij})\right]^2 =
    \begin{cases}
        \left(1 - \cfrac{r_{ij}}{r_c}\right)^2, & r_{ij} \leq r_c \\
        0, & r_{ij} > r_c \\
    \end{cases}
\end{equation}
The values of the parameters of the DPD forces used in this work are reported in \Cref{tab:dpd_simple_fluid_parameters}, together with the number density of beads $\rho$, the temperature $T$, the Boltzmann constant $k_\mathrm{B}$, and the mass of the beads $m$.
These values are consistent with those adopted in the seminal works on DPD \cite{espanol1995statistical, groot1997dissipative}.

\begin{table}[htpb]
    \caption{\label{tab:dpd_simple_fluid_parameters}Parameters used for the DPD simple fluid model.
    All the parameters are expressed in reduced DPD units.}
    \begin{tabular}{cccccccc}
        \hline
        $a$ & $\gamma$ & $\sigma$ & $r_c$ & $\rho$ & $T$ & $k_\mathrm{B}$ & $m$\\
        \hline
        25.0 & 4.5 & 3.0 & 1.0 & 3.0 & 1.0 & 1.0 & 1.0\\
        \hline
    \end{tabular}
\end{table}

\subsection{Non-equilibrium simulations}
\label{subsec:non_equilibrium_simulations}

In atomistic simulations, the shear rheology of a fluid is usually evaluated through non-equilibrium simulations using Lees-Edwards boundary conditions (LEBC) \cite{lees1972computer}.
With this approach, a linear velocity profile is generated and maintained in the simulation box due to the periodicity of the boundary conditions.
The method was developed for molecular dynamics simulations, but has been applied to DPD in previous works \cite{boromand2015viscosity,droghetti2018dissipative,hendrikse2023anionic}.
All simulations in this work were performed using the open-source software LAMMPS (Large-scale Atomic/Molecular Massively Parallel Simulator) \cite{thompson2022lammps}, where LEBC are not implemented in their original formulation.
The alternative approach uses lagrangian rhomboid boundary conditions (LRBC) and is based on the actual deformation of the box in LAMMPS, applied with the \texttt{fix deform} command.
From a theoretical point of view, LEBC and LRBC are equivalent, provided that the velocity profile is taken into account in the periodicity of the boundary conditions \cite{todd2017nonequilibrium}.
The use of these boundary conditions is sufficient to generate a linear velocity profile, and it is referred to as the boundary-driven approach.
Nevertheless, in addition to the lagrangian rhomboid boundary conditions, the SLLOD equations of motion (EoM) are used.
For a planar shear flow applied in the $xy$ plane and a velocity profile in the $x$ direction, the SLLOD EoM reduce to the following form \cite{todd2017nonequilibrium}:
\begin{equation}\label{eq:sllod_equation_of_motion}
    \begin{aligned}
        \dot{\boldsymbol{r}}_i &= \frac{\boldsymbol{p}_i}{m_i} + \boldsymbol{\mathrm{i}}\dot{\gamma} y_i \\
        \dot{\boldsymbol{p}}_i &= \sum \boldsymbol{f}_{i} - \boldsymbol{\mathrm{i}}\dot{\gamma} p_{yi},
    \end{aligned}
\end{equation}
where the dot superscript indicates a time derivative, except for $\dot{\gamma}$ that is the shear rate, $\boldsymbol{\mathrm{i}}$ is the unit vector in the $x$ direction, $\sum\boldsymbol{f}_{i}$ is the sum of the forces acting on the bead $i$, $\boldsymbol{r}_i$ is the position vector, $\boldsymbol{p}_i$ the momentum, and $m_i$ the mass of the bead $i$.
The use of SLLOD has important advantages with respect to the boundary-driven approach.
It provides a direct link with response theory and the possibility of studying time-dependent flows \cite{evans1990statistical,todd2017nonequilibrium}.

In LAMMPS, the SLLOD equations of motion are implemented together with the Nos\'e-Hoover thermostat \cite{maffioli2024ttcf4lammps}:
\begin{equation}\label{eq:lammps_sllod}
    \begin{aligned}
        \dot{\boldsymbol{r}}_i &= \frac{\boldsymbol{p}_i}{m_i} + \boldsymbol{\mathrm{i}}\dot{\gamma} y_i \\
        \dot{\boldsymbol{p}}_i &= \sum \boldsymbol{f}_{i} - \boldsymbol{\mathrm{i}}\dot{\gamma} p_{yi} - \alpha\dot{\boldsymbol{p}}_i \\
        \dot{\alpha} &= \frac{1}{Q}\left(\sum_i \boldsymbol{p}_i^2 - 3 N k_B T\right),
    \end{aligned}
\end{equation}
where $\alpha$ is the multiplier of the Nos\'e-Hoover thermostat, $Q$ is the damping parameter, and $N$ is the number of particles in the system.
While the presence of a thermostat is necessary for a Lennard-Jones model in order to obtain an NVT ensemble, it is not needed for DPD simulations.
As previously described, the dissipative and random forces of the DPD model act as a built-in thermostat.
A potential interaction between the DPD and the Nos\'e-Hoover thermostat is not trivial to evaluate and could lead to incorrect results.
Moreover, integrating the additional equation of motion for the thermostat multiplier increases the computational cost of the simulations.
To avoid these issues, it is preferable to deactivate the Nos\'e-Hoover thermostat while performing DPD simulations.
A quick fix for this purpose is to set the \texttt{tdamp} parameter in LAMMPS to a huge value, \textit{e.g.} $10^{30}$.
The parameter \texttt{tdamp} is related to the relaxation time of the temperature, so the thermostat becomes more aggressive when this value becomes smaller.
By setting it to a very large value, the relaxation time is larger than the length of the simulation, effectively deactivating the thermostat.

The alternative pursued in this work is to modify the LAMMPS source code to create a new fix that applies the SLLOD equations of motion without the thermostat.
The new fix is called \texttt{nve/sllod} and it is available at the following repository link: \url{https://github.com/f2a-dr/sllod_nve}.
It is important to emphasize that, despite the name of the fix, when it is used together with a DPD model, the resulting ensemble is an NVT. 

After the system setup, during a non-equilibrium simulation, it is possible to calculate the apparent viscosity $\mu$ of the fluid using Newton's law of viscosity:
\begin{equation}\label{eq:newton_viscosity}
    \mu = -\frac{\langle P_{yx}\rangle}{\dot{\gamma}},
\end{equation}
where the shear rate $\dot{\gamma}$ is an input parameter of the simulation, proportional to streaming velocity and the deformation of the box, and $P_{yx}$ is the shear pressure.
In this case, the off-diagonal term of the pressure tensor of interest is $P_{yx}$ because the gradient of the $x$-component of the velocity is non-zero along the $y$ direction.
The elements of pressure tensor are evaluated using the 
Virial \citep{parker1954tensor} formula:
\begin{equation}\label{eq:virial}
    P_{\alpha\beta} = \frac{1}{V}\sum_{i=1}^N \left[ m_i v_{i,\alpha} v_{i,\beta} + \frac{1}{2}\sum_{j \ne i}^{N} r_{ij,\alpha}F_{ij,\beta} \right],
\end{equation}
where $V$ is the volume of the simulation box, the subscripts $\alpha$ and $\beta$ refer to the Cartesian components, and the subscripts $i$ and $j$ refer to different beads.
As reported in \Cref{eq:newton_viscosity}, we are interested in the off-diagonal component $P_{yx}$, but, in LAMMPS (version 29Aug2024), the pressure tensor is evaluated considering it as always symmetric, hence $P_{yx} = P_{xy}$.

As previously mentioned, identifying appropriate conversion factors is crucial for translating the results of DPD simulations into meaningful physical units of measurement.
For systems in equilibrium, these factors can be directly calculated from the characteristics of the fluid modeled.
Hence, if a single DPD bead represents a water molecule, a conversion factor for the length can be calculated from the approximate volume of the molecule (\textit{i.e.} $\approx 30\ \text{\AA}^3$).
This value can be associated with the sphere of radius equal to the cutoff radius $r_c$, resulting in a length conversion factor on the order of $10^{-10}\ \mathrm{m}$.
Analogously, the mass of a water molecule, which is approximately $3\times 10^{-26}\ \mathrm{kg}$ can be compared to the unitary mass of a bead, yielding a mass conversion factor equal to the mass of a water molecule.
Various approaches can be used to derive a conversion factor for time.
A possibility is to match the value of $k_\mathrm{B}T$ in real units with the DPD one, which is typically set to unity, to calculate the conversion factor for energy.
The next step is to use the three obtained conversion factors to derive the time conversion factor.
An alternative approach is based on matching the real value of the self-diffusion coefficient of water with the one from DPD simulations \cite{groot2001mesoscopic}.
In both cases, the resulting conversion factor for time is typically on the order of $10^{-12}\ \mathrm{s}$.
The shear rate conversion factor is the reciprocal of the one for time and is therefore on the order of $10^{12}\ \mathrm{s}^{-1}$.
With these conversion factors, a shear rate of $\dot{\gamma} = 0.01$ DPD units provides an acceptable signal-to-noise ratio (SNR) but translates to approximately $10^{10}\ \mathrm{s}^{-1}$ in real units.
Such shear rates are not representative of conditions in industrial equipment, nor they are accessible experimentally.
In rheometry experiments, the range of shear rate is usually between $10^{-1} - 10^{3}\ \mathrm{s}^{-1}$, which corresponds to $10^{-13} - 10^{¨-9}$ in DPD units.
Therefore, performing simulations at lower shear rates with a high SNR is essential for studying the fluid under more realistic conditions.

\subsection{Transient Time Correlation Function}
\label{subsec:transient_time_correlation_function}

To calculate the apparent viscosity as in \Cref{eq:newton_viscosity}, the value of $P_{yx}$ is averaged over many realizations, or trajectories, of the same simulations.
This simple averaging procedure is the most common approach for calculating the shear viscosity in atomistic simulations, and it will be referred to as direct averaging (DAV) throughout this work.
The most important drawback of DAV is linked to the inherent noise of the simulations, which can completely conceal the signal of interest, especially for small shear rates.
To avoid this issue and increase the signal-to-noise ratio (SNR), high shear rates are usually applied to the system. 
This approach guarantees a high SNR and is suitable for Newtonian fluids, but it poses some complications when applied to non-Newtonian fluids.
In many cases, the shear rate applied to achieve high SNR results is too high to be compared with experimental data or even realistic industrial applications.
If the viscosity depends on the shear rate, as for non-Newtonian fluids, the results may not be representative of the real system, and extrapolation to lower shear rates may not be accurate.

An alternative approach to DAV is the transient-time correlation function (TTCF), which is a generalization of the Green-Kubo relations \cite{evans1990statistical,evans2016fundamentals}, and it states that \cite{evans1990statistical,evans2016fundamentals}:
\begin{equation}\label{eq:general_ttcf}
    \langle B(t) \rangle = \langle B(0) \rangle + \int_0^t \langle \Omega(0) B(s) \rangle \mathrm{d}s,
\end{equation}
where $B(t)$ is a generic phase variable measured in the system and $\Omega(0)$ is the dissipation function evaluated at time $t=0$, the instant at which the external field driving the system out of equilibrium is applied.
The dissipation function is related to the external dissipative field applied to the system, and to the work done by this field.
In the case of a sheared system with SLLOD dynamics, the dissipation function is equal to:
\begin{equation}\label{eq:sllod_dissipation_function}
    \Omega = - \frac{\dot{\gamma}V}{k_B T}P_{yx}.
\end{equation}
Consequently, the equation for the evaluation of the shear pressure becomes:
\begin{equation}\label{eq:pyx_sllod_ttcf}
    \langle P_{yx}(t) \rangle = \langle P_{yx}(0) \rangle - \frac{\dot{\gamma}V}{k_B T} \int_0^t \langle P_{yx}(0) P_{yx}(s) \rangle \mathrm{d}s.
\end{equation}

As shown in \Cref{eq:general_ttcf}, the TTCF formalism correlates $\Omega(0)$, a quantity computed at the equilibrium, with $B(t)$, which is obtained from the non-equilibrium trajectories, $B(t)$.
In practice, the simulation procedure is based on a single equilibrium trajectory, called ``mother'', which is used to spawn many non-equilibrium trajectories, called ``daughters''.
In this way, the initial conditions for the non-equilibrium trajectories are generated from the equilibrium probability distribution of the system \cite{todd2017nonequilibrium,maffioli2024ttcf4lammps}.
The mother trajectory is then sampled to be used as the daughters' initial condition at regular intervals, which must be long enough to ensure that the starting points of different daughters are decorrelated.
Moreover, decorrelation of the quantities in the non-equilibrium trajectories is a condition for the use of TTCF, which means:
\begin{equation}\label{eq:decorrelation}
    \langle \Omega(0) B(t) \rangle \rightarrow \langle \Omega(0) \rangle \langle B(t) \rangle, \quad \text{for } t \rightarrow \infty.
\end{equation}
Under this condition, the system is \textit{mixing} and the convergence of the integral is ensured.
At $t=0$ the system is in equilibrium, hence the dissipation function $\langle \Omega(0) \rangle = 0$, and after the decorrelation time the integral does not contribute anymore to $B(t)$.
Nonetheless, from a computational point of view, $\langle \Omega(0) \rangle$ is equal to zero only in the limit of an infinite number of trajectories.
With a finite number of trajectories, the integrated function will not go to zero after the decorrelation time, and the integral value will continue to grow indefinitely in time.

The most common approach to ensure that $\langle \Omega(0) \rangle = 0$ is to generate ensemble members with the same probability but different non-equilibrium trajectories from the same point in the equilibrium trajectory \cite{evans1990statistical}.
To do so, the positions and momenta of the equilibrium space phase point $\boldsymbol{\Gamma}_i$ are modified according to mappings that depend on the type of external field applied to the system.
In the case of SLLOD with planar shear flow in the $xy$-plane, the following mappings are a potential choice:
\begin{equation}\label{eq:sllod_mappings}
    \begin{aligned}
        \boldsymbol{\Gamma}_i &= (\boldsymbol{x},\boldsymbol{y},\boldsymbol{z},\boldsymbol{p}_x,\boldsymbol{p}_y,\boldsymbol{p}_z) \\
        \boldsymbol{\Gamma}_i^{'} &= (\boldsymbol{x},\boldsymbol{y},\boldsymbol{z},-\boldsymbol{p}_x,-\boldsymbol{p}_y,-\boldsymbol{p}_z) \\
        \boldsymbol{\Gamma}_i^{''} &= (\boldsymbol{x},-\boldsymbol{y},\boldsymbol{z},\boldsymbol{p}_x,-\boldsymbol{p}_y,\boldsymbol{p}_z) \\
        \boldsymbol{\Gamma}_i^{'''} &= (\boldsymbol{x},-\boldsymbol{y},\boldsymbol{z},-\boldsymbol{p}_x,\boldsymbol{p}_y,-\boldsymbol{p}_z).
    \end{aligned}
\end{equation}
Choosing the mappings in this way means that $P_{yx}(0) = P_{yx}^{'}(0) = -P_{yx}^{''}(0) = -P_{yx}^{'''}(0)$, hence $\langle P_{yx}(0) \rangle = 0$.
Moreover, the mappings increase the efficiency of the simulations, as they allow the generation of multiple daughter trajectories from a single sample of the mother trajectory.
For these reasons, most of the literature cited in the present work successfully employs the TTCF together with mappings.

\subsubsection{The use of mappings with a DPD model}
\label{subsubsec:dpd_and_mappings}

The main role of mappings is to ensure that the dissipation function is equal to zero at the time the external field is applied.
To check whether this effect is maintained when using a DPD model, the shear pressure $P_{yx}$ is calculated using the Virial formula \Cref{eq:virial}.
From \Cref{eq:sllod_dissipation_function}, the only time-dependent variable in the definition of the dissipation function for the SLLOD is $P_{yx}$, hence $P_{yx}(0) = 0 \Rightarrow \Omega(0) =0$.

The first sum in \Cref{eq:virial} is the kinetic term, which depends only on the velocities of the particles.
Consequently, it is not directly affected by the functional form of the force field used in the simulation.
Applying the mappings will then lead to a zero contribution of the kinetic term for a DPD model, as is the case for a Lennard-Jones fluid.
The second sum is the configurational term, which depends on the forces acting on the particles.
It is possible to show that the contribution of the conservative force $\boldsymbol{F}^C$ is equal to zero when mappings are applied, since it depends only on the position of the beads.
The random force $\boldsymbol{F}^R$ follows the same argument only if the random number $\xi_{ij}$ is the same for the interaction between the beads $i$ and $j$ in all the mapped trajectories.
A different result is obtained for the dissipative force $\boldsymbol{F}^D$, which depends on the relative velocity between the beads.
In this case, the sum across the mappings of the contribution of this force to the configurational term is equal to zero only in equilibrium simulations, \textit{i.e.} when no velocity profile is imposed on the box.
When the external field is applied and the velocity profile is imposed, the following result is obtained:
\begin{equation}\label{eq:dissipative_force_mappings}
    \sum_\text{mappings} r_{ij,x} F^D_{ij,y} = -4 \gamma \dot{\gamma} w^D(r_{ij}) \left(\frac{y_i - y_j}{r_{ij}}\right) x_i (x_i - x_j),
\end{equation}
which is equal to zero only if the two beads $i$ and $j$ have the same $x$ or $y$ coordinates.
In particular, from \Cref{eq:dissipative_force_mappings}, it is clear that this contribution is dependent on the imposed shear rate $\dot{\gamma}$.

\begin{figure}[H]
    \centering
    \includegraphics{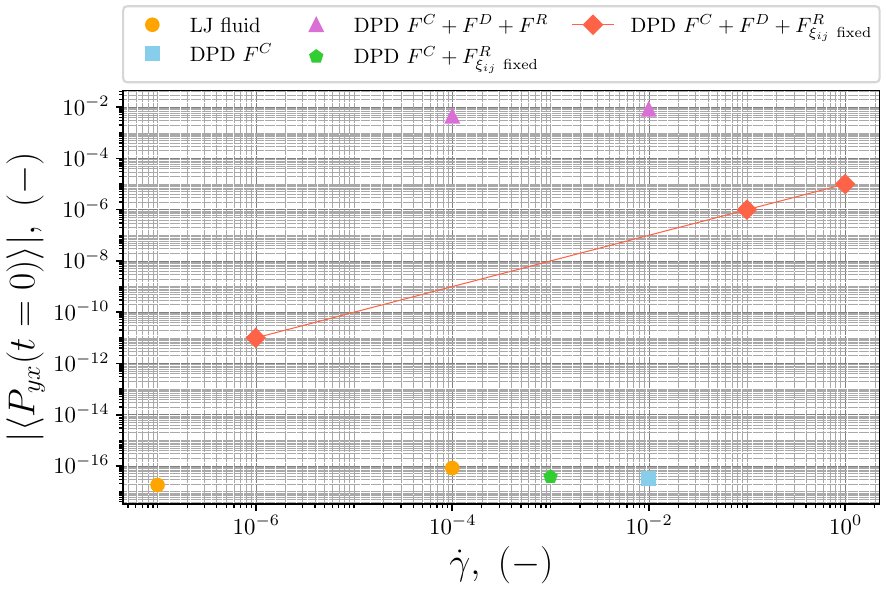}
    \caption{Values of $\lvert \langle P_{yx}(0)\rangle\rvert$ for different models using mappings.
    The yellow circles refer to the LJ-WCA fluid, the purple triangles to a standard DPD model, the light blue square to a DPD model with only conservative force, the green pentagons to a DPD model without dissipative force and with a constant number $\xi$ for the random force, and the red diamonds to a DPD model with the three standard forces but a constant random number $\xi$.
    $\langle P_{yx}(0) \rangle$ can assume negative values, so its absolute value is plotted to use a logarithmic scale.}
    \label{fig:omega0_different_models}
\end{figure}

The results just described are derived in more detail in \Cref{app:mappings_with_dpd_force_field}, and have been tested using simulations of Lennard-Jones and DPD models.
These tests are illustrated in \Cref{fig:omega0_different_models}, where the values of $\lvert \langle P_{yx}(0)\rangle\rvert$ are reported for different force fields.
The mapping should ensure that the shear stress is zero at time $t=0$, which is seen in the simulation results as $\lvert \langle P_{yx}(0)\rangle\rvert \sim O(10^{-16})$, a value of zero given the limits of the machine precision (double).
The results of the Lennard-Jones WCA model are used as a benchmark and confirm the expected behavior of the mappings with a value of zero to machine precision.
On the contrary, for a standard DPD model (``DPD $F^C + F^D + F^R$'' in the plot), the initial value of the shear stress is considerably different from zero.
This is a result of the random terms, which are unique for each daughter and so do not cancel across mappings.
Therefore, using different random numbers for all daughters masks the expected dependence on the shear rate reported in \Cref{eq:dissipative_force_mappings}.
The DPD model using only the conservative force, indicated with ``DPD $F^C$'' in the figure, shows a value of zero.
To further explore the influence of the random number $\xi$, a modified DPD model was tested in this part of the work.
The model ``DPD $F^C + F^R_{\xi_{ij}\ \mathrm{fixed}}$'' in \Cref{fig:omega0_different_models} uses only the conservative and random force, but the value of $\xi$ is set equal for every random number $\xi_{ij}$, regardless of the beads involved in the interaction. 
This shows the same cancellation property across mappings and zero initial shear stress.
Eventually, the approach of a fixed constant random number is used with the complete DPD force field (``DPD $F^C + +F^D + F^R_{\xi_{ij}\ \mathrm{fixed}}$'' in the plot), where the initial shear stress increases together with the shear rate $\dot{\gamma}$.
The simulations confirm the theoretical findings, including the dependence of the dissipative contribution of the configurational term on the shear rate.

As a consequence, it is not possible to use the mappings in \Cref{eq:sllod_mappings} together with a DPD model to ensure that $\langle P_{yx}(0) \rangle = 0$.
In principle, different mappings could be developed to respect the condition also for the dissipative force, but forcing the values of $\xi_{ij}$ to be constant across the mapped trajectories poses a different problem.
Enforcing this condition would require non-trivial management of the random numbers, since the interaction between bead $i$ and bead $j$ must use the same random number $\xi_{ij}$ for each mapped trajectory.
Setting the same random seed in each mapped daughter will not be sufficient, as the order of operations would also need to be identical.
A potential solution could involve defining the same list of random numbers for each bead in each mapped trajectory and deriving $\xi_{ij}$ from the numbers $\xi_i$ and $\xi_j$ in the list.
Such an implementation would be complex in parallel simulations, with the added difficulty of avoiding unexpected correlations.

The alternative to the use of mappings is a modification of the TTCF formula in \Cref{eq:pyx_sllod_ttcf}, to take into account the finite number of trajectories.
\citeauthor{hartkamp2012transienttime} developed this modification and applied it to a LJ-WCA fluid under mixed and elongational flow, also providing an intuitive explanation for the modification.
Considering $\langle \Omega(0) \rangle \neq 0$ as an error, this error can be subtracted from $\Omega(0)$ in the integrand function, so it is possible to write:
\begin{equation}\label{eq:sllod_correction}
    \Big\langle B(s) \Big[ \Omega(0) - \Big\langle \Omega(0) \Big\rangle \Big] \Big\rangle =  \Big\langle B(s) \Omega(0) - B(s) \Big\langle \Omega(0) \Big\rangle \Big\rangle = \Big\langle B(s) \Omega(0) \Big\rangle - \Big\langle B(s) \Big\rangle \Big\langle \Omega(0) \Big\rangle,
\end{equation}
which is equal to the covariance between $B(s)$ and $\Omega(0)$.\\
Consequently, the modified formula for the shear pressure becomes:
\begin{equation}\label{eq:pyx_sllod_ttcf_modification}
    \begin{split}
        \Big\langle P_{yx}(t) \Big\rangle &= \Big\langle P_{yx}(0) \Big\rangle - \frac{\dot{\gamma}V}{k_B T} \int_0^t \Big[\Big\langle P_{yx}(0) P_{yx}(s) \Big\rangle - \Big\langle P_{yx}(0) \Big\rangle \Big\langle P_{yx}(s) \Big\rangle \Big] \mathrm{d}s \\
        &= \Big\langle P_{yx}(0) \Big\rangle - \frac{\dot{\gamma}V}{k_B T} \int_0^t \Big\langle P_{yx}(0) P_{yx}(s) \Big\rangle \mathrm{d}s + \frac{\dot{\gamma}V}{k_B T} \Big\langle P_{yx}(0) \Big\rangle \int_0^t \Big\langle P_{yx}(s) \Big\rangle \mathrm{d}s.
    \end{split}
\end{equation}

\subsubsection{Error estimation}
\label{subsubsec:error_estimation}

The main advantage of TTCF over DAV is the high signal-to-noise ratio that can be obtained from simulations even at very low shear rates.
On the other hand, the DAV is a more straightforward method, easy to implement and to use, while the TTCF requires a more complex setup.
In this context, error estimation becomes a crucial point in the choice of the method to calculate the shear viscosity.
When using the formulation in \Cref{eq:pyx_sllod_ttcf}, the error estimation is simple and direct, since the variance of the left-hand side of the equation is equal to the sum of the variances of the two terms on the right-hand side.
When using the TTCF without mappings, \Cref{eq:pyx_sllod_ttcf_modification} can be rearranged since the ensemble averages and integrals are linear operators:
\begin{equation}\label{}
    \langle P_{yx}(t) \rangle = \langle P_{yx}(0) \rangle - \frac{\dot{\gamma}V}{k_B T} \left\langle \int_0^t P_{yx}(0) P_{yx}(s) \mathrm{d}s \right\rangle + \frac{\dot{\gamma}V}{k_B T} \langle P_{yx}(0) \rangle \left\langle \int_0^t P_{yx}(s) \mathrm{d}s \right\rangle.
\end{equation}
From this equation, it is clear that the variance of $\langle P_{yx}(t) \rangle$ cannot be calculated as the sum of the variances of the three averaged variables.
This is due to the product of the two ensemble averages $\langle P_{yx}(0) \rangle \left\langle \int_0^t P_{yx}(s) \mathrm{d}s \right\rangle$ in the last term.

Since a different approach is required to estimate the precision of TTCF, the bootstrap method \cite{good1999resampling} is used in this work.
This method allows the recontruction of an approximation of the distribution of an estimator by resampling the dataset.
It was described for the first time by \citeauthor{efron1979bootstrap} and employs resampling with replacement, which means that the same value can be sampled multiple times.
In the case of TTCF, the collected data set is large enough, but the particular formulation makes the calculation of the variance impossible.
In this case, it is possible to use the bootstrap method to estimate the distribution of the mean, and from this compute its 95\% confidence interval.

For each time step, a number of samples equal to the number of trajectories is drawn from the ensemble of $P_{yx}$ values obtained from the simulations.
The mean is then calculated from the sampled dataset and the whole resampling procedure is repeated a number of times decided by the user.
A distribution of the mean is then obtained and the 95\% confidence interval is calculated from this distribution.
Moreover, an estimation of the standard error of the mean can be obtained from the same distribution.
A higher number of trajectories or resamples will produce more accurate results, but it can also dramatically increase the computational cost of the boostrapping procedure.

\subsection{Computational details}
\label{subsec:computational_details}

The TTCF formalism is implemented in the open-source Python package TTCF4LAMMPS \cite{maffioli2024ttcf4lammps} built on top of LAMMPS and available at \url{https://github.com/edwardsmith999/TTCF4LAMMPS}.
The original package was modified to include the possibility of using the TTCF without mappings and a module to perform bootstrapping on the generated data has been added.
With respect to the original version, the approach without mappings requires the user to save the variable of interest, \textit{i.e.} $P_{yx}$, for every time step of every trajectory.
As a result, both the simulation process and the bootstrapping procedure are embarrassingly parallelizable, but the disk space required to store the data can be very large.
In particular, it is proportional to the number of trajectories, the number of timesteps per trajectory, and the number of variables of interest.

For all the results presented in this work, the simulations of the Lennard-Jones WCA fluid have been performed on a system of $N = 256$ particles, with a time step of $\Delta t = 2.5 \times 10^{-3}$ in reduced units.
The initial mother trajectory is run for 10000 timesteps to ensure the system reaches equilibrium, then sampled every 1000 timesteps to generate a total of 40000 daughter trajectories, each running for 600 timesteps.

The DPD simulations presented here were performed in a box of side $L = 5$ DPD reduced units, corresponding to a total of $N = 375$ particles, considering the number density $\rho$, the conservative force coefficient $a$, the dissipative force coefficient $\gamma$, the random force coefficient $\sigma$, and the cut-off radius $r_c$ reported in \Cref{tab:dpd_simple_fluid_parameters}.
The timestep is set to $\Delta t = 0.01$ DPD reduced units, the mother trajectory is initially equilibrated for 1500 timesteps and sampled every 100 timesteps to generate a total of $1\times 10^5$ daughter trajectories, each running the sheared system for 420 timesteps.
The criteria for the choice of simulation length and timestep value are illustrated respectively in \Cref{app:identifying_simulation_length} and \Cref{app:influcence_of_timestep_on_dpd_simulations}.
The bootstrapping was performed only on the DPD results by resampling the original dataset 1200 timer, with a sample size equal to the number of daughter trajectories.

\section{Results and discussion}
\label{sec:results_and_discussion}

\subsection{Reproduction of LJ results with and without Mappings}
\label{subsec:lj_no_mappings}

Before using the TTCF non-mapped approach with a DPD system, its performances are assessed on an LJ-WCA fluid.
Previous works studied the application of TTCF on a simple LJ-WCA fluid to compute the shear viscosity, providing a benchmark.
In particular, the setup used by Maffioli \textit{et al.} \cite{maffioli2024ttcf4lammps} is reproduced here.
To better understand the plot of this section, it is useful to recall that the expected viscosity for this fluid at the simulated shear rates is $\mu \sim 2.3\textrm{ - }2.4$ (--) \cite{todd2017nonequilibrium,pan2006prediction,maffioli2024ttcf4lammps}.
Consequently, the expected value of shear pressure is $P_{yx} \sim 2.3\cdot\dot{\gamma}$ (--).

\begin{figure}[H]
    \centering
    \includegraphics[scale=0.85]{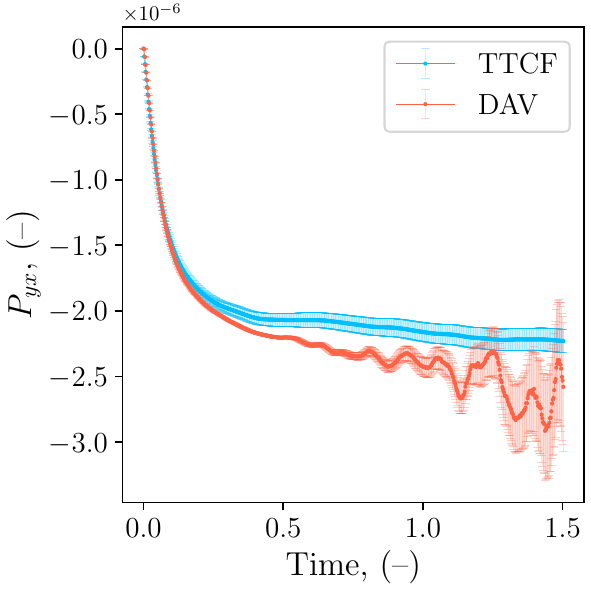}
    \caption{Time response of $P_{yx}$ for a LJ-WCA fluid with an applied shear rate of $\dot{\gamma} = 10^{-6}\ (\text{--})$.
    Four mappings are used, the error bars are equal to the standard error.}
    \label{fig:lj_pyx_mappings}
\end{figure}

\Cref{fig:lj_pyx_mappings} illustrates the typical results obtained with a direct ensemble average of the trajectories (DAV) compared to the TTCF ones.
The use of mapping is necessary to ensure that $\langle\Omega(0)\rangle$ is equal to zero, which for a simple shear means $\langle P_{yx}(0)\rangle = 0$ (\textit{cf.} \Cref{eq:sllod_dissipation_function}).
Additionally, the mappings lead to a reduction of the DAV's standard error in the initial phase of the non-equilibrium simulation.
Hence, the DAV's error, which is zero initially due to the mappings, grows over the simulation as each daughter trajectory diverges.
As $t\to \infty$, the DAV's standard error will reach a maximum value, which is related to the effective accuracy of the method.
In a similar manner, it is possible to observe in the plot an increase in the TTCF's standard error, which is associated with the integration process.
This means that the error will continue to grow in time without reaching a plateau, making long simulations less precise.
The plot in \Cref{fig:lj_pyx_mappings} clearly shows a higher standard error for DAV at the steady state, when compared to TTCF at such low shear rates.

\begin{figure}[H]
    \centering
    \includegraphics[]{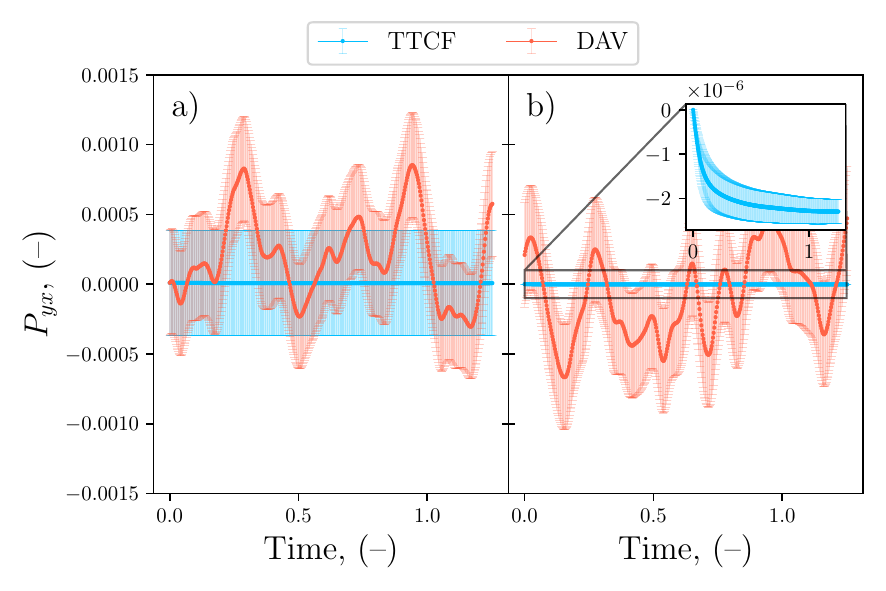}
    \caption{Time response of $P_{yx}$ for a LJ-WCA fluid with an applied shear rate of $\dot{\gamma} = 10^{-6}$ (--).
    Mappings are not used, and the correction in \Cref{eq:pyx_sllod_ttcf_modification} is adopted.
    a) The value of $\langle P_{yx}(0) \rangle$ is calculated as an ensemble average.
    b) The value of $\langle P_{yx}(0) \rangle$ is imposed equal to zero for the TTCF formula.}
    \label{fig:lj_pyx_no_mapping_full}
\end{figure}

When mappings are not used, as in \Cref{fig:lj_pyx_no_mapping_full}a, the high standard error of both DAV and TTCF is visible starting from the first timestep.
A comparison of the two curves shows that the mean value of the DAV oscillates significantly more than that of the TTCF.
The apparently similar standard error in \Cref{fig:lj_pyx_no_mapping_full}a is misleading, as the TTCF error is in practice orders of magnitude lower, but this is the result of the error at time zero introduced by the term $\langle B(0)\rangle$ in \Cref{eq:general_ttcf}, which dominates the plot.
This term is still calculated as a direct ensemble average, and, consequently, it has a standard error of the same order of magnitude as the other DAV measurements.
To eliminate this effect, the equilibrium condition of the mother trajectory is exploited.
For a system in equilibrium $\langle P_{yx}\rangle = 0$, and at $t = 0$ the shear is applied on an equilibrium system, therefore $\langle P_{yx}(0)\rangle = 0$ is imposed in \Cref{eq:pyx_sllod_ttcf_modification}.
Such a procedure is applicable only when the value for a system at equilibrium is known from theory, as in the present case.
Imposing this condition makes the TTCF signal unaffected by the DAV noise in $t = 0$, as shown in \Cref{fig:lj_pyx_no_mapping_full}b.
Moreover, in \Cref{fig:lj_pyx_no_mapping_full}b, it is possible to compare the accuracy of DAV and TTCF for low shear rates by looking at the error bars associated with the two methods.
The DAV standard error is about three orders of magnitude bigger than the signal, while the TTCF error bars indicate a much lower uncertainty, making evident the higher precision of the TTCF.

As already shown in the literature \cite{maffioli2024ttcf4lammps}, the advantage of the TTCF reduces when the shear rate is increased.
At high shear rates, the precision of the DAV becomes comparable and even higher than that of the TTCF one.

\begin{figure}[H]
    \centering
    \includegraphics[]{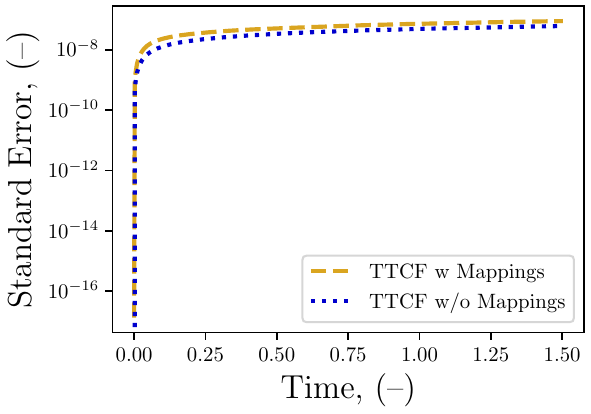}
    \caption{Comparison of the standard error of $P_{yx}$ obtained with and without the mappings using the TTCF on $4\times 10^4$ daughter trajectories.
    The shear rate applied on the LJ-WCA fluid is $\dot{\gamma} = 10^{-6}$ (--).}
    \label{fig:map_vs_nomap_se}
\end{figure}

The use of mappings is generally considered beneficial as the canceling of errors between daughter trajectories, starting from the same point in phase space, is used to reduce uncertainty \citep{evans1988transienttimecorrelation}.
As a result, removing these mappings, as required by the DPD, would be expected to perform worse than the mapped approach.
The error was tested by applying a low shear rate of $10^{-6}$ and the comparison between the mapped and non-mapped approaches is presented in \Cref{fig:map_vs_nomap_se}.
Surprisingly, the non-mapped approach does not exhibit a significant increase in error, and, in the tested case, even shows a small reduction.
This implies the benefits of the TTCF are mainly derived from the ensemble of trajectories and not the use of mappings.
Nonetheless, mappings remain useful, as they can generate multiple starting points without advancing the mother trajectory.
For large systems or long correlation times, this can represent a significant computational saving since many more mappings than the four used here can be generated from a single point in time.

\subsection{DPD system}
\label{subsec:dpd_no_mappings}

\subsubsection{Existing data for viscosity of DPD simple fluids}
\label{subsubsec:previous_dpd_viscosity_data}

Many works in the literature are focused on the use of dissipative particle dynamics (DPD) to study complex fluids, whereas interest in simulating simple fluids is limited.
This is most likely due to the remarkable capability of DPD to reproduce the structural properties of complex fluids and the lack of interest in simple fluids from an applicative point of view.
In this work, as the methodology for applying TTCF with DPD is developed, the focus is placed on the simplest DPD fluid.
Among the few works that studied the transport properties of a simple DPD fluid, there is a certain degree of disagreement about the exact value of the viscosity.
\Cref{tab:dpd_viscosity_literature} recaps some of those viscosity values, which have been obtained using both equilibrium and non-equilibrium methods \cite{lauriello2021simulation, panoukidou2021comparison}.
Given the variability obtained by previous studies, these values are intended to set and build the context, rather than as a reference for the simulations.

\begin{table}[htpb]
    \caption{\label{tab:dpd_viscosity_literature}Viscosity valued for a DPD simple fluid from literature.}
    \centering
    \begin{tabular}{ccc}
        \hline
        Reference & Method & Viscosity $\mu$ (--) \\
        \hline
        Lauriello \textit{et al.} \cite{lauriello2021simulation} & Green-Kubo & 0.860 \\
        Lauriello \textit{et al.} \cite{lauriello2021simulation} & Einstein-Helfand & 0.847 \\
        Panoukidou \textit{et al.} \cite{panoukidou2021comparison} & Green-Kubo & 1.1 \\
        Panoukidou \textit{et al.} \cite{panoukidou2021comparison} & Lees-Edwards & 1.1 \\
        Droghetti \textit{et al.} \cite{droghetti2018dissipative} & Lees-Edwards & 0.85 \\
        Boromand \textit{et al.} \cite{boromand2015viscosity} & Green-Kubo & 0.86 \\
        Boromand \textit{et al.} \cite{boromand2015viscosity} & Lees-Edwards & 0.97 \\
        \hline
    \end{tabular}
\end{table}

\subsubsection{DPD without mappings}
\label{subsubsec:dpd_no_mappings}

As presented in \Cref{fig:omega0_different_models}, due to the presence of dissipative and random forces, the use of mappings is not enough to guarantee that $\langle\Omega(0)\rangle = 0$ for DPD systems.
To consider this issue, the correction described in \Cref{subsubsec:dpd_and_mappings} is applied to the TTCF formula, and the value of $\langle P_{yx}(0)\rangle$ is imposed to be zero.
Finally, as previously noted, the original LAMMPS implementation of SLLOD includes a Nos\'e-Hoover thermostat, which can interfere with the built-in thermostat of DPD.
As described in \Cref{subsec:non_equilibrium_simulations}, this issue can be overcome in two ways: either by modifying the LAMMPS source code, or by using \texttt{nvt/sllod} with a very long thermostat relaxation time (\textit{i.e.} $\texttt{t\_damp} = 10^{30}$).
Both methods yield the same results, but those presented in this section were obtained using the modified version of SLLOD, \texttt{nve/sllod}, which does not apply the Nos\'e-Hoover thermostat.

\begin{figure}[H]
    \centering
    \includegraphics[width=\textwidth]{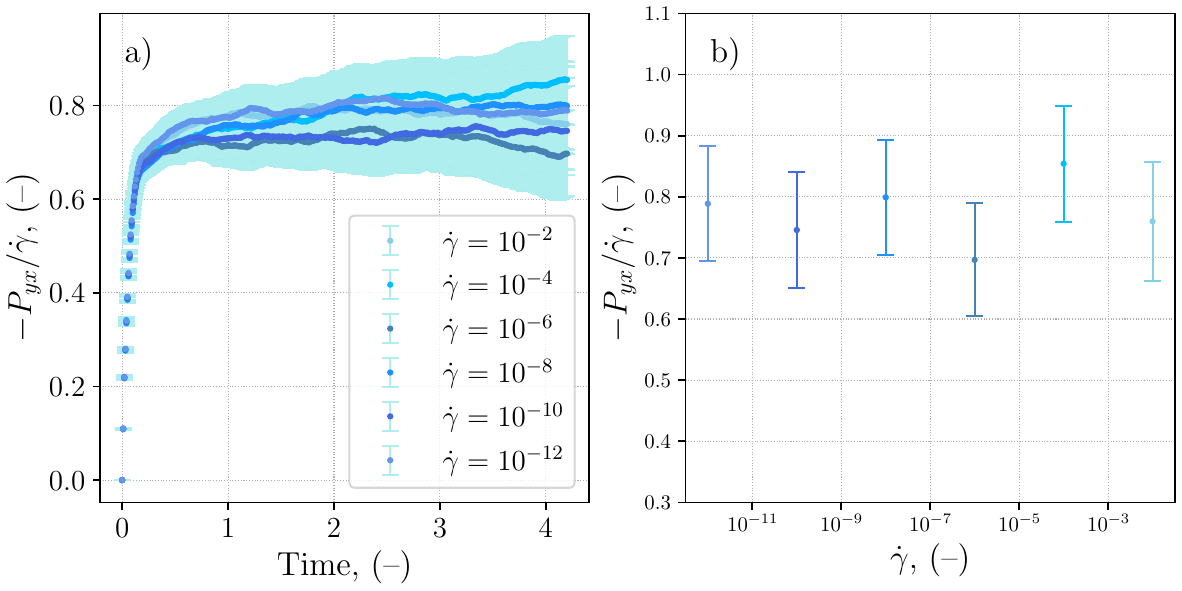}
    \caption{Results obtained with the TTCF method on a DPD simple fluid: a) time evolution of the shear pressure $P_{yx}$ divided by the shear rate $\dot{\gamma}$ for different shear rates; b) value for different shear rates at the last timestep of the simulation.
    The error bars represent the 95\% confidence interval for the mean value.
    All values are expressed in reduced DPD units.}
    \label{fig:mu_ttcf_nve}
\end{figure}

The DPD system is investigated at different shear rates, ranging from $10^{-12}$ to $10^{-2}$ (reduced DPD units).
The results of the simulations are summarized in \Cref{fig:mu_ttcf_nve}, which shows the value of the shear pressure $P_{yx}$ divided by the shear rate $\dot{\gamma}$.
Without this normalization, the values of $P_{yx}$ would differ by several orders of magnitude, depending on $\dot{\gamma}$, making the comparison impossible.
Moreover, $-P_{yx}/\dot{\gamma}$ is equal to the viscosity, which is expected to be constant since a simple DPD fluid exhibits Newtonian behavior.
Looking at the time evolution in \Cref{fig:mu_ttcf_nve}, it is clear that the same viscosity value is reached after the transient for all studied shear rates.
These curves are plotted with different shades of blue to highlight the absence of any visible trend with respect to the shear rate.
The error bars reported in both plots of \Cref{fig:mu_ttcf_nve} correspond to the 95\% confidence interval for the mean value, and they are used to assess the precision of the method.
As expected, the error in \Cref{fig:mu_ttcf_nve}a increases with time, due to the error accumulated during the numerical evaluation of the integrals.
This is an inherent error of the method, which can be limited either by using a more accurate integration algorithm or by reducing the number of timesteps used in the simulation.
Hence, identifying the end of the transient is crucial to obtaining the lowest possible uncertainty.

\begin{figure}[H]
    \centering
    \includegraphics[width=\textwidth]{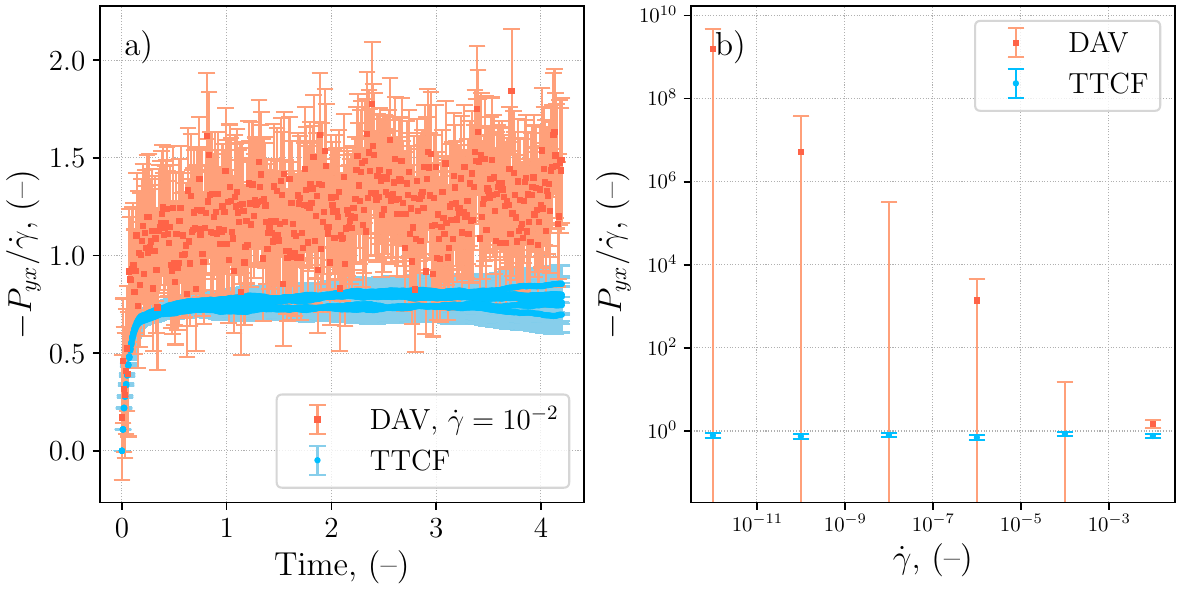}
    \caption{Comparison between DAV and TTCF for a DPD simple fluid.
    The error bars represent the 95\% confidence interval for the mean value.
    a) Time evolution of the shear pressure $P_{yx}$ divided by the shear rate $\dot{\gamma}$, DAV results are shown only for a shear rate of $10^{-2}$ (--).
    b) Value obtained at the last timestep of the simulation for different shear rates, the DAV error bars are not symmetrical due to the logarithmic scale of the $y$-axis.
    All the values are in reduced DPD units.}
    \label{fig:mu_dav_ttcf_nve}
\end{figure}

The error bars in \Cref{fig:mu_ttcf_nve}b are the most important result obtained with the TTCF, as they show how the precision of the method is not affected by the value of the shear rate.
For comparison, \Cref{fig:mu_dav_ttcf_nve} includes the results obtained with the DAV method, which is the standard approach to compute the shear viscosity in DPD simulations.
From \Cref{fig:mu_dav_ttcf_nve}b, it is possible to understand how DAV precision and accuracy are affected by the shear rate.
When the shear rate is reduced, the DAV error bars grow dramatically, making it impossible to use the results obtained from the simulation for $\dot{\gamma} < 10^{-2}$.
For the same reason, in \Cref{fig:mu_dav_ttcf_nve}a the DAV results are shown only for the highest shear rate tested, $\dot{\gamma} = 10^{-2}$.
Plotting the curves for lower shear rates would render the figure unreadable, due to the excessive noise present in DAV results.
Moreover, from this plot, it is possible to notice a certain discrepancy between the DAV and TTCF time evolution.
This behavior was already observed in previous works that used LAMMPS \cite{maffioli2024ttcf4lammps}, and it is most likely due to issues in the implementation of SLLOD in LAMMPS.
The same discrepancy is noticeable in the simulations of the LJ-WCA fluid in \Cref{fig:lj_pyx_mappings}, and leads to a systematically higher value of the viscosity calculated with the DAV.
Currently, this issue is being addressed by other research groups, which are working on an alternative implementation of the SLLOD algorithm in LAMMPS \cite{ssande7lammps}.

The increment in precision obtained through the use of TTCF is assessed more in detail using two quantities.
The first is the standard error (SE), the second is a measure of the relative magnitude between the signal and the noise, indicated as signal-to-error ratio (SER).
The SER is caclulated as:
\begin{equation}
    \mathrm{SER} = \frac{\langle P_{yx} \rangle}{\mathrm{SE}_{P_{yx}}},
\end{equation}
where the ensemble average of the shear pressure is divided by its standard error.
Since the DAV presents too high uncertainty for low shear rates, the SER is calculated using the mean value from the TTCF method, leading to the following expressions:
\begin{equation}
    \mathrm{SER^{DAV}} = \frac{\langle P_{yx} \rangle^{TTCF}}{\mathrm{SE}_{P_{yx}}^{DAV}};\qquad
    \mathrm{SER^{TTCF}} = \frac{\langle P_{yx} \rangle^{TTCF}}{\mathrm{SE}_{P_{yx}}^{TTCF}};
\end{equation}

\begin{figure}[H]
    \centering
    \includegraphics[width=\textwidth]{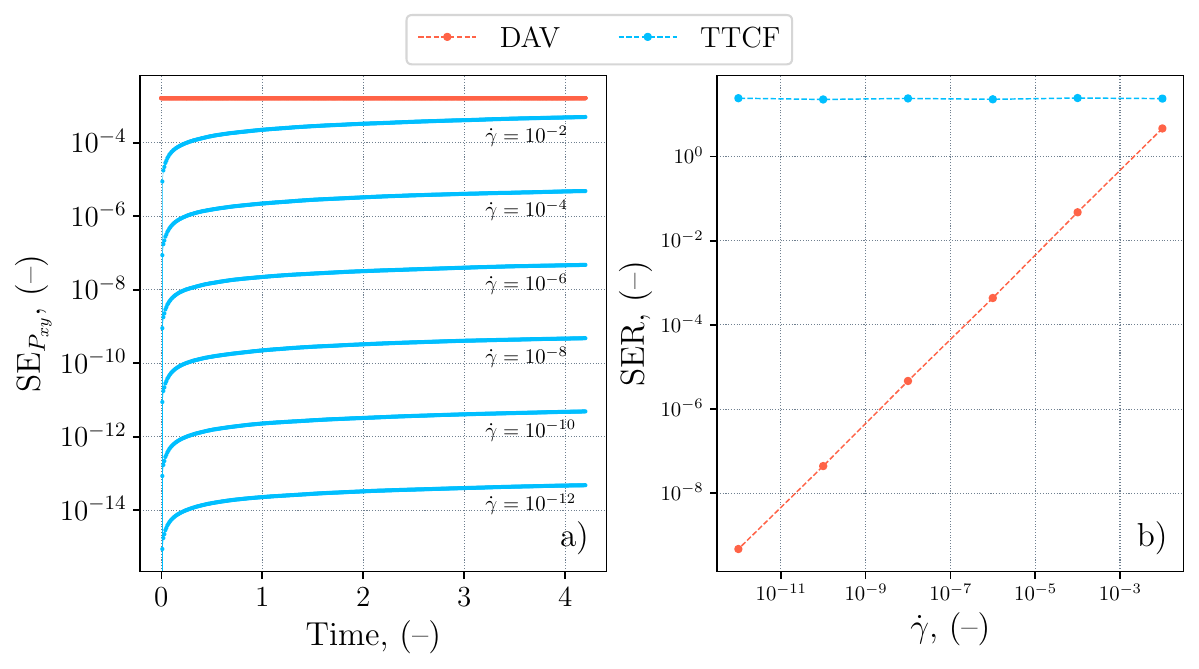}
    \caption{Precision assessment of DAV and TTCF for a DPD simple fluid.
    a) Time evolution of the standard error (SE) for the shear pressure for different shear rates. The DAV curves collapse on a single one, while each TTCF curve refers to a different shear rate.
    b) Signal-to-noise ratio (SNR) for the shear pressure calculated on the last timestep, the mean value from TTCF is used as signal for both TTCF and DAV curves.
    All values are expressed in reduced DPD units.}
    \label{fig:se_snr_nve}
\end{figure}

The results are plotted in \Cref{fig:se_snr_nve}, and show a quantitative comparison between the precision of the two methods.
The DAV standard error in \Cref{fig:se_snr_nve}a is constant in time, as expected, since mappings are not used.
Moreover, the DAV curves collapse on a single one, indicating a standard error that does not depend on the shear rate.
The limitations of the DAV are evident, as its performance is inferior to that of TTCF in modeling $\dot{\gamma} \leq 10^{-2}$.
In contrast, the TTCF standard error grows in time as a result of the numerical integration, but has the advantage of being proportional to the shear rate.
This confirms the TTCF as a suitable method for arbitrarily low shear rates in DPD simulations, since it means that the SNR is constant with respect to the shear rate.
The plot \Cref{fig:se_snr_nve}b illustrates this behavior, also showing that the SNR for DAV decreases by several orders of magnitude when the shear rate is lowered.

\subsubsection{Computational cost and accuracy}
\label{subsubsec:computational_cost_and_accuracy}

The use of TTCF requires a large number of independent simulations, since it is based on the evaluation of the transient.
On the other hand, each simulation does not usually require a large number of timesteps, depending on the decorrelation time of $\Omega(0)$ and $B(t)$.
As reported in \Cref{subsec:computational_details}, $10^{5}$ trajectories have been simulated and averaged on 420 timesteps for each value of the shear rate.
The management of such a high number of simulations was possible using the Python package TTCF4LAMMPS, modified to include the non-mapped approach.
The simulations were performed on a cluster, using 120 cores over approximately half an hour of wall time, equal to about 63 core hours, for each shear rate.
The bootstrap method was parallelized on the same number of cores, and the time necessary to evaluate the standard error was negligible compared to the simulation time, as reported in \Cref{tab:dpd_simulation_time}.

\begin{table}[htpb]
    \caption{\label{tab:dpd_simulation_time}Computational time to perform simulations and bootstrap for different shear rates.}
    \centering
    \begin{tabular}{ccc}
        \hline
        $\dot{\gamma}$ (--) & Simulation time (s) & Bootstrap time (s)\\
        \hline
        $10^{-2}$ & 1850 & 48\\
        $10^{-4}$ & 1861 & 48\\
        $10^{-6}$ & 1898 & 46\\
        $10^{-8}$ & 1890 & 46\\
        $10^{-10}$ & 1859 & 45\\
        $10^{-12}$ & 1880 & 47\\
        \hline
    \end{tabular}
\end{table}

The study of a DPD simple fluid has some advantages from the computational point of view, since in such systems the decorrelation time is quite short.
This is due to the absence of microstructures, which, depending on the type of structure, can increase the time for the stresses in the fluid to relax.
For this reason, the simulations were carried out on a small system of 375 particles, which resulted in a relatively high uncertainty for both TTCF and DAV.
This is noticeable both in \Cref{fig:mu_ttcf_nve}b and in \Cref{fig:se_snr_nve}b, where the confidence interval for the mean value is still quite large and the TTCF SNR is below one.
The present work is focused on illustrating how to apply the TTCF method to DPD systems and to highlight the advantages with respect to the DAV.
Hence, the absolute value of SNR is not the main focus, but rather the fact that it is constant with respect to the shear rate.

It is possible to increase the accuracy of the TTCF method by either $i)$ considering a large system or $ii)$ increasing the number of independent trajectories to be averaged.
The first approach is often necessary for studying complex fluids, where the size of the microstructures imposes the size of the box.
This must be large enough to be representative of the system.
Increasing the system size is computationally more expensive, but usually increases the SNR.
The second approach allows for an increase in the computational cost and precision of the method with more granularity.
However, in some cases, a large number of independent trajectories is necessary to significantly improve the SNR.

\section{Conlusions}
\label{sec:conclusions}

This work presents the application of the transient time correlation function (TTCF) method to compute the shear viscosity of a simple DPD fluid with non-equilibrium simulations.
It is shown that the TTCF can be successfully applied to DPD systems and illustrated how the computational method must be modified to account for the DPD force field.
The presence of dissipative and random forces of DPD required a modification of the SLLOD algorithm implemented in LAMMPS, to avoid interference of the Nos\'e-Hoover thermostat with the DPD one.
Moreover, it was demonstrated that these two forces, the core of the DPD thermostat, break the symmetry imposed by the mappings.
Mappings are no longer sufficient to guarantee that the dissipation function $\langle\Omega(0)\rangle$ is equal to zero at the initial time, and a correction to the TTCF equation becomes necessary.
The revised formulation, already proposed by previous works, allows the calculation of the apparent viscosity for a DPD simple fluid, but makes it more complex to estimate the error.
To overcome this issue, the bootstrap method was used to estimate the distribution of the mean value, and recover from it the 95\% confidence interval.
The increase in computational cost due to the bootstrapping procedure is negligible compared to the simulation time due to the high parallelizability of the method.
The results of the TTCF method were compared with the direct ensemble average (DAV), which is the standard approach for this kind of simulation.
As expected, the DAV method showed an SNR that decreases with lower shear rates, making the error too high for the DAV to be useful at shear rates below $10^{-2}$ (--).
On the other hand, the TTCF method was able to provide accurate results at any shear rate with a lower error than the DAV and an SNR that remains constant with respect to the shear rate.
The more straightforward DAV method is only reliable for high shear rates, which usually do not correspond to replicable conditions in experiments.
Finally, a key outcome of this study is that the use of mappings is not practical with DPD.
Remarkably, their absence does not compromise the accuracy of TTCF results and can even lead to improved precision.
The results presented open the possibility to study the rheology of structured fluids using DPD and TTCF, where the use of low shear rates to match the experimental conditions would be a crucial improvement.
In this way, the conversion factors could be recovered from the system characteristics, leading to simulations that match the experimental conditions.
Furthermore, the deformation of the microstructures under shear flow could be studied under realistic conditions, avoiding unphysical behavior due to extremely high shear rates.
Studies on such complex systems will also be helpful in identifying the limitations of the method, which are often influenced by system-specific properties, such as the stress relaxation time.

\begin{acknowledgement}

We thank Stephen Sanderson for the insightful discussions about the SLLOD implementation in LAMMPS.
We thank Daniele Dini, Billy Todd, and Peter Daivis for the fruitful discussions about TTCF and its implementation without mappings.
We thank Daniele Marchisio for the general guidance and the feedback on this work.
We are grateful to the UK Materials and Molecular Modelling Hub for computational resources, which is partially funded by EPSRC (EP/T022213/1, EP/W032260/1 and EP/P020194/1).
We acknowledge ISCRA for awarding this project access to the LEONARDO supercomputer, owned by the EuroHPC Joint Undertaking, hosted by CINECA (Italy) .

\end{acknowledgement}

\bibliography{achemso-demo}

\appendix

\newpage 
\section{Mappings with the DPD force field}
\label{app:mappings_with_dpd_force_field}

As reported in \Cref{subsec:transient_time_correlation_function}, the dissipation function at time zero, $\langle \Omega(0) \rangle$, must be equal to zero to use the TTCF in non-equilibrium simulations.
For a system with simple shear applied in the $xy$ plane, $\Omega$ is described by \Cref{eq:sllod_dissipation_function}, here repeated:
\begin{equation}\label{eq:sllod_dissipation_function_app}
    \Omega = - \frac{\dot{\gamma}V}{k_BT}P_{xy}.
\end{equation}
Consequently, to have $\langle\Omega(0)\rangle = 0$, it must be true that $\langle P_{xy}(0) \rangle = 0$.
This condition is verified for a physical system, since $t=0$ corresponds to an equilibrium condition, but it is practically impossible to obtain by averaging a finite number of samples.
In Lennard-Jones systems, this issue can be resolved by using the following mappings:
\begin{equation}\label{eq:mappings_complete}
\begin{alignedat}{2}
    & \boldsymbol{\Gamma}_i && = (\boldsymbol{x}, \boldsymbol{y},  \boldsymbol{z}, \boldsymbol{p}_x, \boldsymbol{p}_y, \boldsymbol{p}_z)\\
    & \boldsymbol{\Gamma}_i^{\mathrm{I}} && = (\boldsymbol{x}, \boldsymbol{y},  \boldsymbol{z}, -\boldsymbol{p}_x, -\boldsymbol{p}_y, -\boldsymbol{p}_z)\\
    & \boldsymbol{\Gamma}_i^{\mathrm{II}} && = (-\boldsymbol{x}, \boldsymbol{y},  \boldsymbol{z}, -\boldsymbol{p}_x, \boldsymbol{p}_y, \boldsymbol{p}_z)\\
    & \boldsymbol{\Gamma}_i^{\mathrm{III}} && = (-\boldsymbol{x}, \boldsymbol{y},  \boldsymbol{z}, \boldsymbol{p}_x, -\boldsymbol{p}_y, -\boldsymbol{p}_z),\\
\end{alignedat}
\end{equation}
which generate values of $P_{yx}(0)$ equal in modulus and opposite in sign for the mapped trajectories, giving zero as a result.
It is possible to show that these mappings are not suitable for DPD systems, due to the presence of the dissipative and random forces.
In LAMMPS, the pressure tensor $\boldsymbol{P}$ is calculated using the Irving-Kirkwood formula and is always considered symmetric.
Hence, the $P_{yx} = P_{xy}$ element of the pressure tensor is computed as \cite{thompson2022lammps}:
\begin{equation}\label{eq:stress_tensor_xy_component}
    P_{xy} = \frac{1}{V}\sum_{k=1}^N m_k v_{kx}v_{ky} + \frac{1}{V}\sum_{k=1}^{N'}r_{kx}f_{ky},
\end{equation}
where the subscript $k$ refers to the $k$-th particle, and $N \neq N'$ due to periodic boundary conditions and communications between processors.
The first sum in \Cref{eq:stress_tensor_xy_component} is the kinetic contribution, while the second is the configurational contribution, or the virial term.
The value of $P_{xy}$ is equal to zero at a certain time if both contributions are equal to zero.
Using the mappings of \Cref{eq:sllod_mappings}, the kinetic terms of the mapped trajectories are:
\begin{equation}\label{eq:kinetic_energy_contribution}
    \begin{alignedat}{1}
        (p_{kx}, p_{ky}, p_{kz}) &\rightarrow \sum_{k=1}^{N}m_kv_{kx}v_{ky}\\
        (-p_{kx}, -p_{ky}, -p_{kz}) &\rightarrow \sum_{k=1}^{N}m_k(-v_{kx})(-v_{ky})= \sum_{k=1}^{N}m_kv_{kx}v_{ky}\\
        (-p_{kx}, p_{ky}, p_{kz}) &\rightarrow \sum_{k=1}^{N}m_k(-v_{kx})v_{ky} = -\sum_{k=1}^{N}m_kv_{kx}v_{ky}\\
        (-p_{kx}, -p_{ky}, -p_{kz}) &\rightarrow \sum_{k=1}^{N}m_kv_{kx}(-v_{ky})= -\sum_{k=1}^{N}m_kv_{kx}v_{ky},\\
    \end{alignedat}
\end{equation}
leading to a zero averaged value for the kinetic contribution to $P_{xy}$.
The contribution of the configurational part can be studied by considering the forces acting on a single bead $k$.
Indeed, for the mappings to work, the sum of the virial terms obtained with the different mappings must be zero for every particle.
To simplify the problem, the following consider the interaction of the bead $k$ with only one bead $l$.

The conservative (\Cref{eq:dpd_conservative_force}) and random (\Cref{eq:dpd_random_force}) forces depend only on the relative position, hence the focus can be put on the transformation of the positions:
\begin{equation}
    \begin{alignedat}{2}\label{eq:mapping_positions}
    & \boldsymbol{\Gamma}_i && = (\boldsymbol{x}, \boldsymbol{y},  \boldsymbol{z})\\
    & \boldsymbol{\Gamma}_i^{\mathrm{I}} && = (\boldsymbol{x}, \boldsymbol{y},  \boldsymbol{z})\\
    & \boldsymbol{\Gamma}_i^{\mathrm{II}} && = (-\boldsymbol{x}, \boldsymbol{y},  \boldsymbol{z})\\
    & \boldsymbol{\Gamma}_i^{\mathrm{III}} && = (-\boldsymbol{x}, \boldsymbol{y},  \boldsymbol{z})\\
\end{alignedat}
\end{equation}

Having only one type of bead, the conservative force between beads $k$ and $l$ is:
\begin{equation}
    \boldsymbol{F}^C_{kl} = aw_C(r_{kl})\hat{r}_{kl}.
\end{equation}
From this equation, $r_{kl}$ and $\hat{r}_{kl}$ can be expressed in an explicit form:
\begin{equation}
\begin{alignedat}{2}
    &\boldsymbol{r}_k &&= (x_k, y_k, z_k),\\
    &\boldsymbol{r}_l &&= (x_l, y_l, z_l),\\
    &\boldsymbol{r}_{kl} &&= (x_k-x_l, y_k-y_l, z_k-z_l) = (x_{kl},y_{kl},z_{kl}),\\
    &r_{kl} &&= |\boldsymbol{r}_{kl}| = \sqrt{x_{kl}^2+y_{kl}^2+z_{kl}^2},\\
    &\hat{r}_{kl} &&= \frac{\boldsymbol{r}_{kl}}{r_{kl}} = \left(\frac{x_k-x_l}{r_{kl}},\frac{y_k-y_l}{r_{kl}},\frac{z_k-z_l}{r_{kl}}\right).\\
\end{alignedat}
\end{equation}
Then, the weight function $w_C$, for $r_{kl}\leq r_c$, is:
\begin{equation}\label{eq:weight_function_explicit}
    w_C(r_{kl}) = \left(1-\frac{r_{kl}}{r_c}\right) = \left(1-\frac{\sqrt{(x_k-x_l)^2+(y_k-y_l)^2+(z_k-z_l)^2}}{r_c}\right).
\end{equation}
When comparing \Cref{eq:weight_function_explicit} with the mapping in \Cref{eq:mapping_positions}, it is possible to notice that the weight function $w_C(r_{kl})$ does not depend on the mappings, since it contains only squared differences.
The results from mappings I and II are identical and equal to the following:
\begin{equation}\label{eq:mappings_12_conservative_force}
    \boldsymbol{F}^{C,\mathrm{I,II}}_{kl} = aw_C(r_{kl})\left(\frac{x_k-x_l}{r_{kl}},\frac{y_k-y_l}{r_{kl}},\frac{z_k-z_l}{r_{kl}}\right).
\end{equation}
In the case of mappings III and IV, the conservative force on bead $k$ due to the interaction with bead $l$ is:
\begin{equation}\label{eq:mappings_34_conservative_force}
    \boldsymbol{F}^{C,\mathrm{III,IV}}_{kl} = aw_C(r_{kl})\left(\frac{-x_k+x_l}{r_{kl}},\frac{y_k-y_l}{r_{kl}},\frac{z_k-z_l}{r_{kl}}\right).
\end{equation}
From \Cref{eq:stress_tensor_xy_component}, the virial term is calculated as the sum of the products of the $x$ component of the position vector times the $y$ component of the force.
The latter force is the same for all the mappings:
\begin{equation}
    F^C_{kl,y} = f^C_{k,y} = aw_C(r_{kl})\left(\frac{y_k-y_l}{r_{kl}}\right),
\end{equation}
and the sum is:
\begin{equation}\label{eq:conservative_virial}
\begin{alignedat}{2}
    \sum_\mathrm{mappings} r_{k,x}f^C_{k,y} &= f^C_{k,y}\sum_\mathrm{mappings} r_{k,x} = f^C_{k,y}(r^\mathrm{I}_{k,x}+r^\mathrm{II}_{k,x}+r^\mathrm{III}_{k,x}+r^\mathrm{IV}_{k,x})\\
    &= f^C_{k,y}(x_k+x_k-x_k-x_k) = 0.
\end{alignedat}
\end{equation}

The same argument can be applied to the random force, which has the same weight function as the conservative force, so that what was said for \Cref{eq:weight_function_explicit} is also valid in this case.
\begin{equation}
    \boldsymbol{F}^R_{kl} = \sigma w_R(r_{kl})\frac{\xi_{kl}}{\sqrt{\Delta t}}\hat{r}_{kl}\label{eq:dpd_random}
\end{equation}
In this case, an additional condition must be imposed on the generation of the random numbers.
In order for the forces to cancel out in the sum, the value of $\xi_{kl}$ must be the same for all the mapped trajectories.
If the generated random number $\xi_{kl}$ is correctly set in all mapped trajectories, the random force has identical values for mapping I and II:
\begin{equation}\label{eq:mappings_12_random_force}
    \boldsymbol{F}^{R,\mathrm{I,II}}_{kl} = \sigma w_R(r_{kl})\frac{\xi_{kl}}{\sqrt{\Delta t}}\left(\frac{x_k-x_l}{r_{kl}},\frac{y_k-y_l}{r_{kl}},\frac{z_k-z_l}{r_{kl}}\right),
\end{equation}
and for mappings III and IV:
\begin{equation}\label{eq:mappings_34_random_force}
    \boldsymbol{F}^{R,\mathrm{III,IV}}_{kl} = \sigma w_R(r_{kl})\frac{\xi_{kl}}{\sqrt{\Delta t}}\left(\frac{-x_k+x_l}{r_{kl}},\frac{y_k-y_l}{r_{kl}},\frac{z_k-z_l}{r_{kl}}\right).
\end{equation}
In addition, the calculation for the virial term is identical to the one in \Cref{eq:conservative_virial}, since the $y$ component of the force is the same for all mappings:
\begin{equation}
    F^R_{kl,y} = f^R_{k,y} = aw_R(r_{kl})\left(\frac{y_k-y_l}{r_{kl}}\right),
\end{equation}
\begin{equation}\label{eq:random_virial}
\begin{alignedat}{2}
    \sum_\mathrm{mappings} r_{k,x}f^R_{k,y} &= f^R_{k,y}\sum_\mathrm{mappings} r_{k,x} = f^R_{k,y}(r^\mathrm{I}_{k,x}+r^\mathrm{II}_{k,x}+r^\mathrm{III}_{k,x}+r^\mathrm{IV}_{k,x})\\
    &= f^C_{k,y}(x_k+x_k-x_k-x_k) = 0.
\end{alignedat}
\end{equation}

Unlike the conservative and random forces, the dissipative force depends on the relative velocity between the two beads:
\begin{equation}
    \boldsymbol{F}^D_{kl} = -\gamma w_D(r_{kl})(\hat{r}_{kl}\cdot\boldsymbol{v}_{kl})\hat{r}_{kl}.
\end{equation}
For this expression, it is possible to show that the mappings work as expected for a system at equilibrium, with no velocity field imposed.
When a shear is applied to the box, the velocity profile breaks the symmetry imposed by the mappings, and the elements in the sum for virial term do not cancel out.
In the nonequilibrium simulation, the velocity $\boldsymbol{v}$ of a bead will be the sum of the streaming velocity $\boldsymbol{U}(y)$, the consequence of the shear imposition, and the peculiar velocity $\boldsymbol{u}$.
For a bead $k$:
\begin{equation}
    \boldsymbol{v}_k = \boldsymbol{u}_k + \boldsymbol{U}_k(y),
\end{equation}
or, in component form:
\begin{equation}
    (v_{k,x},v_{k,y},v_{k,z}) = (u_{k,x},u_{k,y},u_{k,z}) + (\dot{\gamma}y_k,0,0) = (u_{k,x}+\dot{\gamma}y_k,u_{k,y},u_{k,z})
\end{equation}
From this definition, the relative velocity $\boldsymbol{v}_{kl}$ between beads $k$ and $l$ can be calculated:
\begin{equation}
\begin{alignedat}{2}
    &\boldsymbol{v}_k &&= (u_{k,x}+\dot{\gamma}y_k, u_{k,y}, u_{k,z}),\\
    &\boldsymbol{v}_l &&= (u_{l,x}+\dot{\gamma}y_l, u_{l,y}, u_{l,z}),\\
    &\boldsymbol{v}_{kl} &&= (u_{k,x}+\dot{\gamma}y_k-u_{l,x}-\dot{\gamma}y_l, u_{k,y}-u_{l,y}, u_{k,z}-u_{l,z})
\end{alignedat}
\end{equation}

As for conservative and random forces, $r_{kl}$ is not affected by the position transform, and the weight function $w_D(r_{kl})$ has the same value for all mappings.
The next step is the evaluation of the dot product $(\hat{r}_{kl}\cdot\boldsymbol{v}_{kl})$, which must be carried out for each mapping.
It should be noted that only the peculiar velocity $\boldsymbol{u}_k$ is affected by the transformations, since the mappings are applied to the equilibrium system before imposing the shear.
In the case of the original equilibrium trajectory, that is, mapping I $(\boldsymbol{x}, \boldsymbol{y}, \boldsymbol{z}, \boldsymbol{p}_x, \boldsymbol{p}_y, \boldsymbol{p}_z)$:
\begin{equation}
\setlength{\jot}{10pt}
\begin{alignedat}{2}
    \hat{r}_{kl}\cdot\boldsymbol{v}_{kl} &=\\
 =& \left(\frac{x_k-x_l}{r_{kl}},\frac{y_k-y_l}{r_{kl}},\frac{z_k-z_l}{r_{kl}}\right)(u_{k,x}+\dot{\gamma}y_k-u_{l,x}-\dot{\gamma}y_l, u_{k,y}-u_{l,y}, u_{k,z}-u_{l,z})\\
    =& \frac{1}{r_{kl}}\left[(x_k-x_l)(u_{k,x}+\dot{\gamma}y_k-u_{l,x}-\dot{\gamma}y_l) + (y_k-y_l)(u_{k,y}-u_{l,y}) + (z_k-z_l)(u_{k,z}-u_{l,z})\right]\\
    =& \frac{1}{r_{kl}}\left[(x_k-x_l)(u_{k,x}-u_{l,x})+\dot{\gamma}(x_k-x_l)(y_k-y_l) + \mathcal{D}_y + \mathcal{D}_z\right]\\
    =& \frac{1}{r_{kl}}(\mathcal{D}_x+\mathcal{D}_y+\mathcal{D}_z) + \frac{\dot{\gamma}(x_k-x_l)(y_k-y_l)}{r_{kl}}\\
    =& \mathcal{D} + \frac{\dot{\gamma}(x_k-x_l)(y_k-y_l)}{r_{kl}} = \mathcal{D} + \mathcal{A}.
\end{alignedat}
\end{equation}
To simplify the notation, the groups $\mathcal{D}_i$, with $i = x,y,z$, $\mathcal{D}$, and $\mathcal{A}$ are introduced:
\begin{equation}
\begin{alignedat}{2}
    &\mathcal{D}_i &&= (i_k - i_l)(u_k - u_l),\\
    &\mathcal{D} &&= \frac{\mathcal{D}_x + \mathcal{D}_y + \mathcal{D}_z}{r_{kl}},\\
    &\mathcal{A} &&= \frac{\dot{\gamma}(x_k-x_l)(y_k-y_l)}{r_{kl}}.\\
\end{alignedat}
\end{equation}
For mapping II $(\boldsymbol{x}, \boldsymbol{y}, \boldsymbol{z}, -\boldsymbol{p}_x, -\boldsymbol{p}_y, -\boldsymbol{p}_z)$:
\begin{equation}
\setlength{\jot}{10pt}
\begin{alignedat}{2}
    \hat{r}_{kl}\cdot\boldsymbol{v}_{kl} &=\\
 =& \left(\frac{x_k-x_l}{r_{kl}},\frac{y_k-y_l}{r_{kl}},\frac{z_k-z_l}{r_{kl}}\right)(-u_{k,x}+\dot{\gamma}y_k+u_{l,x}-\dot{\gamma}y_l, -u_{k,y}+u_{l,y}, -u_{k,z}+u_{l,z})\\
    =& \frac{1}{r_{kl}}(-\mathcal{D}_x-\mathcal{D}_y-\mathcal{D}_z) + \frac{\dot{\gamma}(x_k-x_l)(y_k-y_l)}{r_{kl}} = -\mathcal{D} + \mathcal{A}
\end{alignedat}
\end{equation}
Appling mapping III $(-\boldsymbol{x}, \boldsymbol{y},  \boldsymbol{z}, -\boldsymbol{p}_x, \boldsymbol{p}_y, \boldsymbol{p}_z)$:
\begin{equation}
\setlength{\jot}{10pt}
\begin{alignedat}{2}
    \hat{r}_{kl}\cdot\boldsymbol{v}_{kl} &=\\
 =& \left(\frac{-x_k+x_l}{r_{kl}},\frac{y_k-y_l}{r_{kl}},\frac{z_k-z_l}{r_{kl}}\right)(-u_{k,x}+\dot{\gamma}y_k+u_{l,x}-\dot{\gamma}y_l, u_{k,y}-u_{l,y}, u_{k,z}-u_{l,z})\\
    =& \frac{1}{r_{kl}}(\mathcal{D}_x+\mathcal{D}_y+\mathcal{D}_z) - \frac{\dot{\gamma}(x_k-x_l)(y_k-y_l)}{r_{kl}} = \mathcal{D} - \mathcal{A}
\end{alignedat}
\end{equation}
With mapping IV $(-\boldsymbol{x}, \boldsymbol{y},  \boldsymbol{z}, \boldsymbol{p}_x, -\boldsymbol{p}_y, -\boldsymbol{p}_z)$:
\begin{equation}
\setlength{\jot}{10pt}
\begin{alignedat}{2}
    \hat{r}_{kl}\cdot\boldsymbol{v}_{kl} &=\\
 =& \left(\frac{-x_k+x_l}{r_{kl}},\frac{y_k-y_l}{r_{kl}},\frac{z_k-z_l}{r_{kl}}\right)(u_{k,x}+\dot{\gamma}y_k-u_{l,x}-\dot{\gamma}y_l, -u_{k,y}+u_{l,y}, -u_{k,z}+u_{l,z})\\
    =& \frac{1}{r_{kl}}(-\mathcal{D}_x-\mathcal{D}_y-\mathcal{D}_z) - \frac{\dot{\gamma}(x_k-x_l)(y_k-y_l)}{r_{kl}} = -\mathcal{D} - \mathcal{A}
\end{alignedat}
\end{equation}
Writing explicitly the dissipative force for every mapping:
\begin{alignat}{3}
    &\boldsymbol{F}^{D,\mathrm{I}}_{jk} &&= -\gamma w_D(r_{kl})(\mathcal{D}+\mathcal{A})\left(\frac{x_k-x_l}{r_{kl}},\frac{y_k-y_l}{r_{kl}},\frac{z_k-z_l}{r_{kl}}\right)\\
    &\boldsymbol{F}^{D,\mathrm{II}}_{jk} &&= -\gamma w_D(r_{kl})(-\mathcal{D}+\mathcal{A})\left(\frac{x_k-x_l}{r_{kl}},\frac{y_k-y_l}{r_{kl}},\frac{z_k-z_l}{r_{kl}}\right)\\
    &\boldsymbol{F}^{D,\mathrm{III}}_{jk} &&= -\gamma w_D(r_{kl})(\mathcal{D}-\mathcal{A})\left(\frac{-x_k+x_l}{r_{kl}},\frac{y_k-y_l}{r_{kl}},\frac{z_k-z_l}{r_{kl}}\right)\\
    &\boldsymbol{F}^{D,\mathrm{IV}}_{jk} &&= -\gamma w_D(r_{kl})(-\mathcal{D}-\mathcal{A})\left(\frac{-x_k+x_l}{r_{kl}},\frac{y_k-y_l}{r_{kl}},\frac{z_k-z_l}{r_{kl}}\right)
\end{alignat}
The calculation of the virial term requires the $y$ component of the force, which is not equal for all mappings, and assumes this form:
\begin{equation}
    F^D_{kl,y} = f^D_{k,y} = -\gamma w_D(r_{kl})(\pm\mathcal{D}\pm\mathcal{A})\left(\frac{y_k-y_l}{r_{kl}}\right).
\end{equation}
To easily identify the terms that cancel out and write the sum in a more compact form, the group $\mathcal{A}$ is refactored and $\mathcal{B}$ is introduced:
\begin{equation}
    \mathcal{A} = \frac{\dot{\gamma}(x_k-x_l)(y_k-y_l)}{r_{kl}} = \dot{\gamma}(x_k-x_l)\frac{(y_k-y_l)}{r_{kl}} = \mathcal{B}\frac{(y_k-y_l)}{r_{kl}},
\end{equation}
The result of the sum for the virial term is:
\begin{equation}\label{eq:dissipative_virial}
\begin{alignedat}{2}
    \sum_\mathrm{mappings} r_{k,x}f^D_{k,y} =& r^\mathrm{I}_{k,x}f^{D,\mathrm{I}}_{k,y}+r^\mathrm{II}_{k,x}f^{D,\mathrm{II}}_{k,y}+r^\mathrm{III}_{k,x}f^{D,\mathrm{III}}_{k,y}+r^\mathrm{IV}_{k,x}f^{D,\mathrm{IV}}_{k,y}\\
    =& x_k\left[-\gamma w^D(r_{kl})(\mathcal{D}+\mathcal{A})\left(\frac{y_k-y_l}{r_{kl}}\right)\right] + x_k\left[-\gamma w^D(r_{kl})(-\mathcal{D}+\mathcal{A})\left(\frac{y_k-y_l}{r_{kl}}\right)\right]+\\
    &- x_k\left[-\gamma w^D(r_{kl})(\mathcal{D}-\mathcal{A})\left(\frac{y_k-y_l}{r_{kl}}\right)\right] - x_k\left[-\gamma w^D(r_{kl})(-\mathcal{D}-\mathcal{A})\left(\frac{y_k-y_l}{r_{kl}}\right)\right]\\
    =& -\gamma w^D(r_{kl})\left(\frac{y_k-y_l}{r_{kl}}\right)^2x_k[(\mathcal{D}+\mathcal{B}) + (-\mathcal{D}+\mathcal{B}) -(\mathcal{D}-\mathcal{B}) -(-\mathcal{D}-\mathcal{B})]\\
    =& -\gamma w^D(r_{kl})\left(\frac{y_k-y_l}{r_{kl}}\right)^2x_k[\mathcal{B}+\mathcal{B}+\mathcal{B}+\mathcal{B}]\\
    =& -\gamma w^D(r_{kl})\left(\frac{y_k-y_l}{r_{kl}}\right)^2x_k4\dot{\gamma}(x_k-x_l)\\
    =& -4\gamma\dot{\gamma}w^D(r_{kl})\left(\frac{y_k-y_l}{r_{kl}}\right)^2x_k(x_k-x_l) \ne 0
\end{alignedat}
\end{equation}
The contribution to the configurational term of the dissipative force is different than zero, therefore the mappings are not sufficient to enforce the condition of $\langle P_{xy}(0) \rangle = 0$.
Moreover, it is clear that the sum in \Cref{eq:dissipative_virial} is proportional to the value of the shear rate $\dot{\gamma}$, supporting the results reported in \Cref{fig:omega0_different_models}.

\newpage 
\section{Identifying simulation length}
\label{app:identifying_simulation_length}

As shown in \Cref{fig:se_snr_nve}a, the standard error of the TTCF method increases in time, so a longer simulation result in lower precision.
Under these conditions, the optimal approach would be to set a simulation time exactly equal to the decorrelation time of the stresses in the system studied.
Oftentimes, it is not possible to identify an exact value for the decorrelation time, and increasing the simulation length seem the safest option.
Once the stress are decorrelated, the integral in \Cref{eq:pyx_sllod_ttcf} will give a null contribution, and the value of $P_{yx}$ will remain constant in time.
In practice, the value of $\langle P_{yx}(0)P_{yx}(t) \rangle$ oscillates around zero after the decorrelation, resulting in small variations of the shear pressure.
This behavior, together with the increase in the TTCF's standard error and computational costs, make undesiderable to use simulations time much longer than the stress decorrelation time.

\begin{figure}
    \centering
    \includegraphics{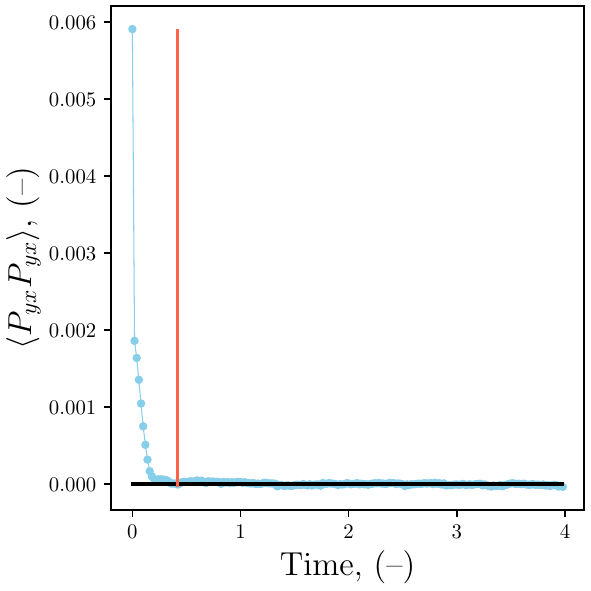}
    \caption{Auto-correlation function for the $P_{yx}$ element of the pressure tensor, calculated from an equilibrium DPD simulation.
    The horizontal black line indicates the value of zero, while the vertical red line identifies the time length of the non-equilibrium simulations performed in this work.}
    \label{fig:pyx_decorrelation}
\end{figure}

An optimal choice of the simulation length probably requires some test in non-equilibrium conditions, to evaluate the best approximation of the stress decorrelation time.
Such workflow could increase dramatically the computational cost, due to the high number of daughter trajectories required.
An alternative approach, used in the present work, consist in evaluating the stress auto-correlation function from an equilibrium simulation to get an initial guess.
The results of this procedure for a simple DPD fluid are reported in \Cref{fig:pyx_decorrelation}, which showed a fast decorrelation and allowed the choice of an appropriate simulation time.
This time value is shown as a vertical red line in \Cref{fig:pyx_decorrelation} and avoided a counterproductively long simulation, even though it cannot be considered an optimal value.

\newpage 
\section{Influence of timestep on DPD simulations}
\label{app:influcence_of_timestep_on_dpd_simulations}

\begin{figure}[H]
    \centering
    \includegraphics{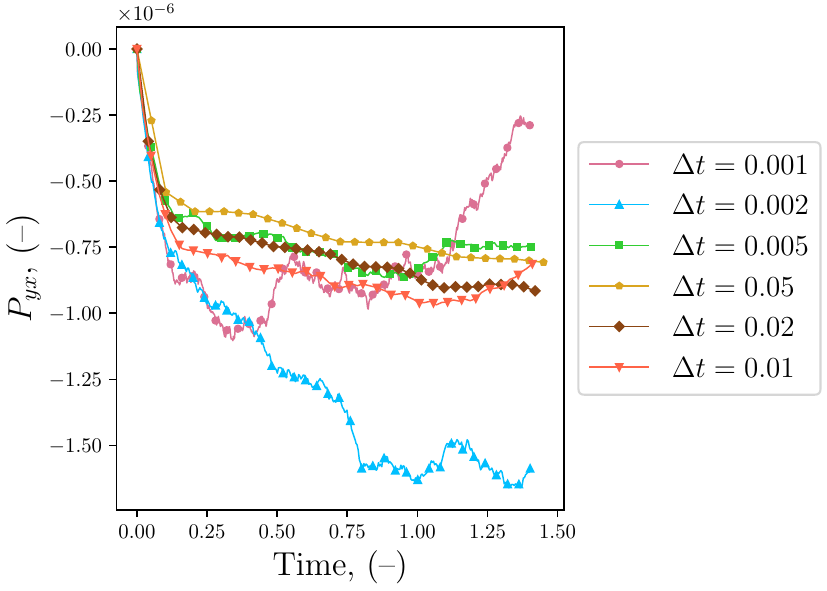}
    \caption{Time evolution of the shear pressure $P_{yx}$ for different values of the timestep $\Delta t$ in DPD simulations.
    All values are obtained using the TTCF method, with a shear rate of $\dot{\gamma} = 10^{-6}$ and $2\times10^{5}$ daughter trajectories.}
    \label{fig:pyx_deltat}
\end{figure}

The value of the timestep $\Delta t$ heavily affect the results of a DPD simulation.
The random force $\boldsymbol{F}^R_{ij}$ depends explicitly on $\Delta t$, as reported in \Cref{eq:dpd_random_force}, differently from many other force fields.
In atomistic simulations, usually a smaller $\Delta t$ allow ot reach a higher accuracy and to capture better the dynamnics of the system.
As explained by \citet{groot1997dissipative}, decreasing the timestep increases the variance of the random force, and, consequently, the noise in the simulation.
An example of this phenomena is reported in figure \Cref{fig:pyx_deltat}, where more noisy and oscillating $P_{yx}(t)$ curves are the result of lower timesteps.

The results in \Cref{fig:pyx_deltat} helped in the choice of the timestep used in this work, which is equal to 0.01.
This is a common value in DPD simulations involving simple fluids and is it the lowest still exhibiting a limited noise, according to the data in the plot.

\end{document}